\documentclass[prb,aps,superscriptaddress,twocolumn]{revtex4-1}

\usepackage{bm}
\usepackage[colorlinks=true,linkcolor=blue,citecolor=blue]{hyperref}
\usepackage{times}
\usepackage{amsmath}
\usepackage{amssymb}
\usepackage{amsthm}
\usepackage{amsfonts}
\usepackage{enumerate}
\usepackage{latexsym}
\usepackage{ifpdf}
\usepackage{natbib}
\newcommand{\beq}{\begin{equation}}
\newcommand{\eeq}{\end{equation}}
\usepackage{graphicx}
\usepackage{makeidx}
\usepackage{color}
\usepackage{array}
\usepackage{multirow}
\usepackage{soul}

\begin{document}

 \title{Conductivity noise across temperature driven transitions of rare-earth nickelate heterostructures}

\author{Gopi Nath Daptary}
\affiliation{Department of Physics, Indian Institute of Science, Bengaluru 560012, India}
\author{Siddharth Kumar}
\affiliation{Department of Physics, Indian Institute of Science, Bengaluru 560012, India}
\author{M. Kareev}
\affiliation{Department of Physics and Astronomy, Rutgers University, Piscataway, New Jersey 08854, USA}
\author{ J. Chakhalian}
\affiliation{Department of Physics and Astronomy, Rutgers University, Piscataway, New Jersey 08854, USA}
\author {Aveek Bid}
\email{aveek@iisc.ac.in}
\affiliation{Department of Physics, Indian Institute of Science, Bengaluru 560012, India}
\author {S. Middey}
\email{smiddey@iisc.ac.in }
\affiliation{Department of Physics, Indian Institute of Science, Bengaluru 560012, India}

\begin{abstract}
The metal-insulator transition (MIT) of bulk  rare-earth nickelates is accompanied by a simultaneous charge ordering (CO) transition. We have investigated low-frequency resistance fluctuations (noise)   across the MIT and magnetic transition of [EuNiO$_3$/LaNiO$_3$] superlattices, where selective suppression of charge ordering has been achieved by mismatching the  superlattice periodicity with the periodicity of charge ordering.  We have observed that irrespective of the presence/absence of long-range CO, the  noise magnitude is enhanced by several orders  with strong non-1/$f$ ($f$ = frequency) component when the system undergoes MIT and magnetic transition.  The higher order statistics of resistance fluctuations reveal the presence of strong non-Gaussian components in both cases, further indicating inhomogeneous electrical transport arising from the electronic phase separation. Specifically, we find  almost three orders of magnitude smaller noise in the insulating phase of the sample without long-range CO compared to the sample  with CO. These findings suggest that digital synthesis can be a potential route to implement electronic transitions of complex oxides for device application.

\end{abstract}

\maketitle

\section{Introduction}

Metal-insulator transition (MIT), observed in complex materials as a function of temperature, chemical doping, electrostatic gating,  magnetic field, light, pressure, epitaxy etc., remains a topic of paramount interest over  decades~\cite{Imada:1998p1039,Tokura:2006p797}. The complexity of the mechanism of MIT in rare earth nickelate series have attracted significant attentions in recent years~\cite{Middey:2016p305,Catalano:2018p046501}. In the bulk form, $RE$NiO$_3$ with $RE$=  Sm, Eu, Lu, Y etc. undergoes a first order transition from an orthorhombic, metallic phase without charge ordering to a monoclinic, insulating phase with a rock-salt type charge ordering (CO)~\cite{Medarde:1997p1679,Catalan:2008p729}.  A magnetic transition (paramagnetic  to $E^\prime$-antiferromagnetic) occurs at a lower temperature.  Moreover,  four transitions appear simultaneously in bulk NdNiO$_3$ and PrNiO$_3$. In order to explain the origin of this peculiar MIT, the importance of structural transition~\cite{Mercy:2017p1667}, electron correlations~\cite{Stewart:2011p176401}, charge ordering~\cite{Staub:2002p126402,Mazin:2007p176406}, distribution of ligand holes~\cite{Barman:1994p8475,Mizokawa2000p11263,Park:2012p156402,johnston:2014p106404,Subedi:2015p075128,Bisogni:2016p13017}, polaron condensation~\cite{Shamblin:2018p86}, Fermi surface nesting~\cite{Lee:2011p016405,Lee:2011p165119,Hepting:2014p227206} etc. have been emphasized by different types of experimental probes and theoretical methods.  Interestingly, it has been demonstrated recently that  a MIT without any long-range CO and   structural symmetry change can be obtained in  the artificial structure of $RE$NiO$_3$ by mismatching the  periodicity of the heterostructure with the periodicity of  rock-salt type CO~\cite{Middey:2018p156801}. Apart from the interest arising from the aspect of fundamental physics, $RE$NiO$_3$ based heterostructures also show excellent potentials for  electronics applications~\cite{nno_fet, Shi2013, Shi2014, Ha2014,Middey:2016p305,Catalano:2018p046501}.

The low-frequency 1/$f$ noise  is not only used for  semiconductor device characterizations~\cite{noisedevice}, but also acts a powerful tool to probe exotic phenomena like electronic phase separation~\cite{koushik2011evidence}, structural phase transition~\cite{chandni2009criticality}, charge density wave~\cite{kundu2017quantum}, superconductor-normal state phase transition~\cite{clarke1976low,babic20071} etc. The frequency dependence of the power spectral density (PSD) $S_R(f)$ (described later in the text) arises due to finite relaxation of the fluctuating variable.  According to the central limit theorem,  the fluctuation statistics of a system is Gaussian if the fluctuators are independent of each other \cite{reif2009fundamentals}. However, the presence of any correlations due to magnetic, electronic, or structural interactions  in the system would result in non-Gaussian statistics of time dependent fluctuations. This information can be extracted from   higher order statistics of resistance fluctuations  via `second spectrum'~\cite{RevModPhys.60.537, ghosh2004set}. The phase transitions of SmNiO$_3$, NdNiO$_3$ single-crystalline films have been studied by such noise and  second spectrum measurements~\cite{ sahoo2014conductivity,alsaqqa2017phase,Bisht:2017p115147}.  The extremely large magnitude of noise and second spectrum have been attributed to the coexistence of metal and insulator phases near the electronic transition temperature.  Such $1/f$ noise study can  also provide crucial information about the length scale of charge ordering as reported earlier for colossal magnetoresistive (CMR)- manganites~\cite{bid2003low}.

 In this work, we  report on resistance fluctuations   across the electronic and magnetic transitions of  [2uc  EuNiO$_3$/1uc LaNiO$_3$] (2ENO/1LNO) and [1uc  EuNiO$_3$/1uc LaNiO$_3$] (1ENO/1LNO) films (uc=unit cell in pseudo-cubic notation).   1ENO/1LNO superlattice (SL) exhibits four simultaneous transitions~\cite{Middey:2018p156801}, similar to bulk NdNiO$_3$ and PrNiO$_3$. On the other hand,  2ENO/1LNO SL  is a rare example, which undergoes  a first-order MIT without any long-range CO and  remains monoclinic in both metallic and insulating phases~\cite{Middey:2018p156801}.   We have observed the random telegraphic noise (RTN) as well as non-Gaussian component  (NGC) of noise near the MIT of these films,  which confirms the coexistence of spatially separated metallic and insulating  phases in both samples.  Importantly, we have found that the energy barrier, separating these electronic phases   and the associated length scale of nanoscopic phase separation are similar in both samples.  However, the noise magnitude in the insulating phase of 2ENO/1LNO SL is {\it three-orders of magnitude smaller} compared to the corresponding noise in 1ENO/1LNO SL, suggesting that the system having MIT without long-range charge ordering would be a better candidate  for practical device applications.  Interestingly, the higher order statistics of resistance fluctuations (quantified as second spectrum) becomes maximum near the antiferromagnetic transition temperature ($T_N$) of 2ENO/1LNO SL, implying certain role of $E^\prime$-magnetic ordering in opening gap in the multi-band Fermi surface.

\section{Experimental details}
[2uc EuNiO$_3$/1uc LaNiO$_3$]x12 (2ENO/1LNO) and [1uc EuNiO$_3$/1uc LaNiO$_3$]x18 (1ENO/1LNO) superlattices (SLs) have been grown on single crystalline NdGaO$_3$ (110) substrate by pulsed laser interval deposition. The details of the growth conditions and  characterizations can be found in Refs.~\onlinecite{Middey:2018p156801,Middey:2018p045115,Middey:2018apl}. The  resistance and noise measurements have been performed in a cryo-free 4 K system.

\section{Results and Discussions}
\begin{figure}
\begin{center}
\includegraphics[width=0.47\textwidth]{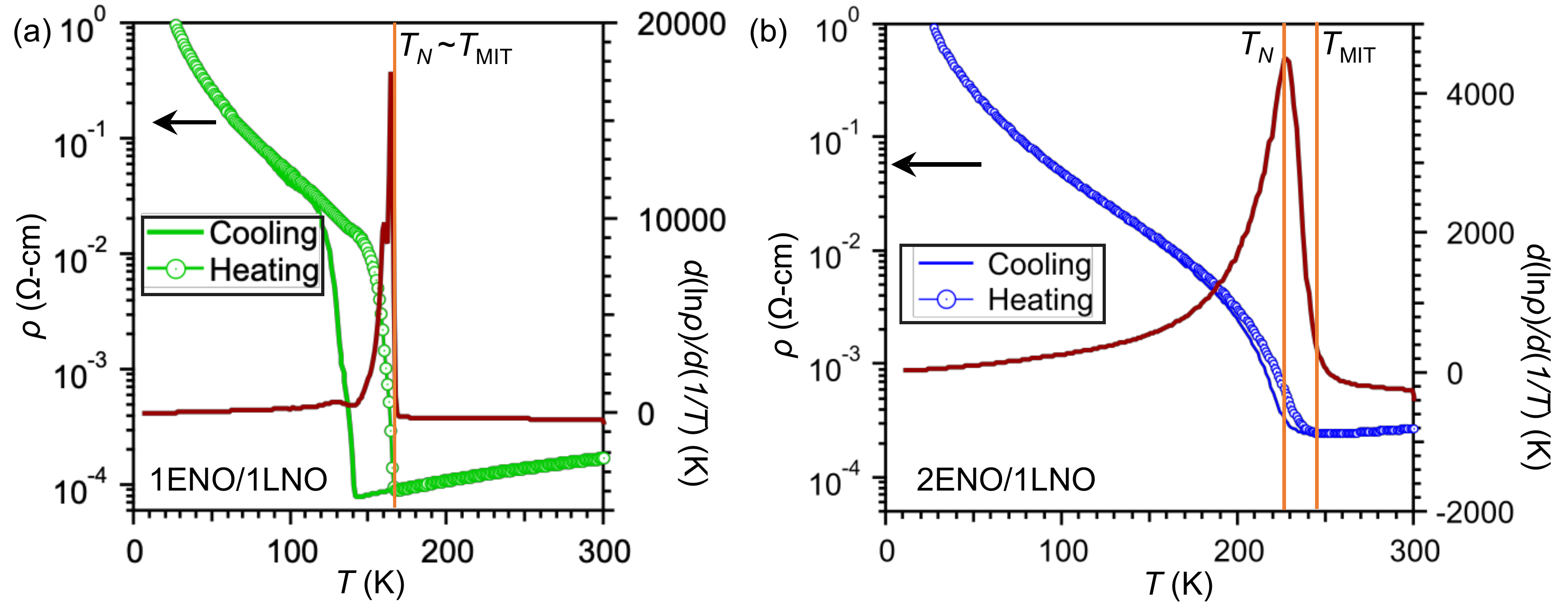} \small{\caption{ Resistivity ($\rho$) as a function of temperature for (a) 1ENO/1LNO and (b) 2ENO/1LNO.  Corresponding  $d \ln (\rho)/d(1/T)$ has been also plotted as a function of $T$ (right axis of (a) and (b)). \label{fig:Fig1}}}
	\end{center}
\end{figure}

Figure \ref{fig:Fig1}(a) and (b) show the temperature dependent resistivity ($\rho$) for 1ENO/1LNO and 2ENO/1LNO films, respectively.  From now onwards, we discuss the results of the heating run.  As seen, 1ENO/1LNO and 2ENO/1LNO SLs undergo first-order insulator to metal transitions around 165 K and 245 K respectively. The magnetic transition temperatures ($T_N$) are found to be 165 K for 1ENO/1LNO and 225 K for 2ENO/1LNO SL from $d \ln (\rho)/d(1/T)$ vs. $T$ plot~\cite{Middey:2018p045115,Zhou:2005p226602,Ojha2019} [see right axis of Fig. \ref{fig:Fig1}(a) and (b)] .

 \begin{figure*}
\begin{center}
\includegraphics[width=0.95\textwidth]{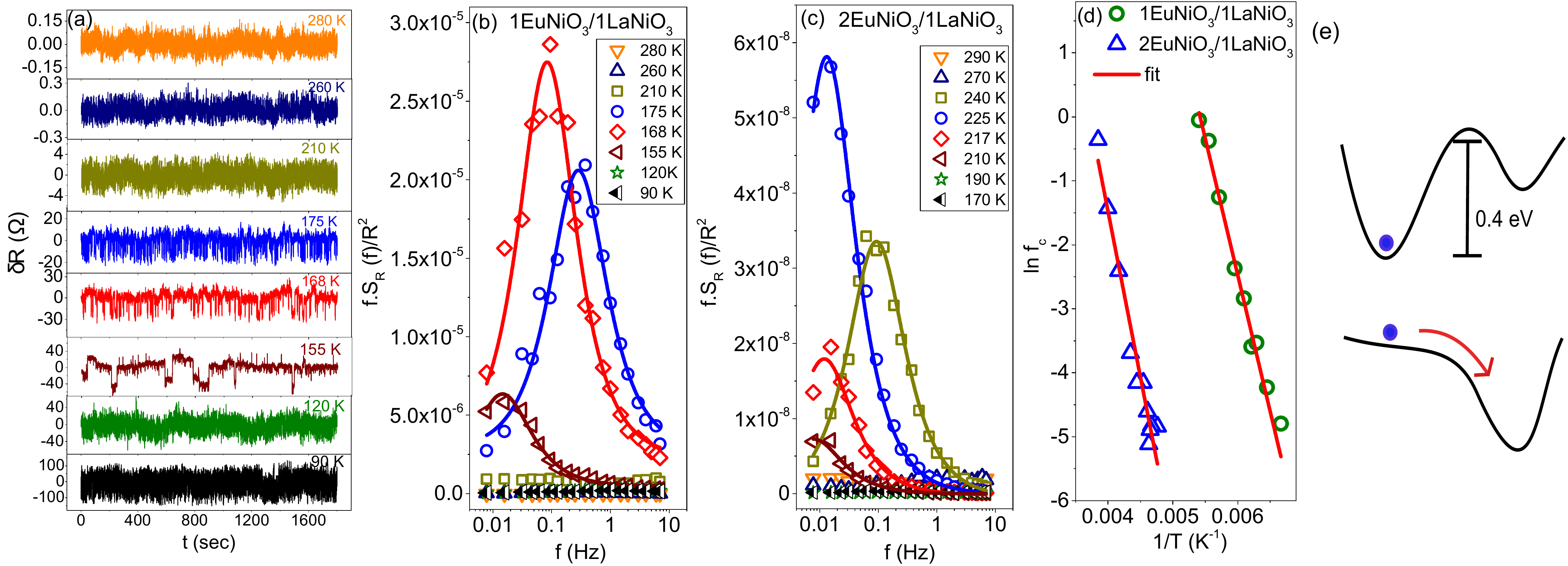}
	\small{\caption{(a) Time series of resistance fluctuations at few representative values of $T$ of 1EuNiO$_3$/1LaNiO$_3$. (b) and (c) Scaled PSD of resistance fluctuations, $f S_R(f)/R^2$ as a function of frequency at a few representative values of $T$ for 1EuNiO$_3$/1LaNiO$_3$ and 2EuNiO$_3$/1LaNiO$_3$ respectively. The solid lines are the fits to the data with Eq. \ref{eqn:lorentzian}. For details, see text. (d) Plot of $f_c$ as a function of inverse temperature with semi-log scale. Solid lines are an Arrhenius fit to the data as discussed in the text. (e) A schematic energy diagram of two level states: insulating and metallic states. Upper and lower panel represent the position of electron (filled circle) while the system is in insulating and metallic state respectively.   \label{fig:Fig2}}}
	\end{center}
\end{figure*}

To probe  nature of the electrical transport, we have measured  low frequency resistance fluctuations of  1ENO/1LNO and 2ENO/1LNO films using standard 4 probe lock-in (LIA) technique~\cite{ghosh2004set}.
This technique allows to measure both the sample as well as background noise. The sample has been current biased ($I$) with an excitation frequency $f^* \sim$ 220 Hz. The voltage fluctuations $\delta V(t)$ arise at the sideband of $f^*$ after the signal is demodulated from the LIA. The output of the LIA has been digitized to a high speed analog to digital converter  (ADC) and stored   to get the time series of voltage fluctuations $\delta V(t)$. The time series of voltage fluctuations $\delta V(t)$ has been converted to time series of resistance fluctuations $\delta R(t)$ as $\delta R(t) = \delta V(t)/I$. In Fig. \ref{fig:Fig2}(a), we plot the time series of resistance fluctuations at different temperatures for 1ENO/1LNO SL. As clearly seen,    each time series  for $T>T_\mathrm{MIT}$ consists of random resistance fluctuations about the average value.  Similar features have been also seen in case of  2ENO/1LNO SL for $T>T_\mathrm{MIT}$ (see Appendix, Fig. \ref{fig:Fig5}).  Interestingly, we observe the appearance of random telegraphic noise (RTN) with the resistance fluctuations between two states  in the temperature range 140 K$<T<200$ K for 1ENO/1LNO SL and  200 K$<T<$ 260 K for 2ENO/1LNO SL. These RTN are absent below 140 K for 1ENO/1LNO SL and  195 K for 2ENO/1LNO SL. Such RTN has been also reported for other systems   e.g. manganites~\cite{bid2003low}, and two-dimensional superconductor~\cite{kundu2017quantum}, where the system can fluctuate between two distinct phases. Surprisingly, the ratio of the temperature ($T_\mathrm{RTN}$), where RTN starts to appear and the  $T_\mathrm{MIT}$ is very similar ($\sim$ 0.85) for both 1ENO/1LNO and 2ENO/1LNO SLs.

To understand the origin of the RTN, we have   investigated the power spectral density (PSD) of the resistance fluctuations $S_R(f)$. At each $T$, the resistance fluctuations have been recorded for 30 minutes. The data have been decimated and digitally filtered to eliminate 50 Hz line frequency. $S_R(f)$ has been calculated using fast Fourier transformation (FFT) technique from the filtered time series \cite{ghosh2004set}. The minimum and maximum frequency of the noise measurement are 4 mHz and 8 Hz, respectively. In order to accentuate any deviation from $1/f$ nature of the spectrum, we have plotted the quantity $fS_R(f)/R^2$  as a function of   $f$ at few representative temperatures for  1ENO/1LNO (Fig. \ref{fig:Fig2}(b)) and 2ENO/1LNO  (Fig. \ref{fig:Fig2}(c)). For $T>>T_{MIT}$, the PSD $S_R(f)$ follows  $1/f^\alpha$ dependence with $\alpha \sim 1$ for both  samples.   However,  a strong deviation from the $1/f$ dependence of the spectral power has been found within the temperature range 140 K$< T< 200$ K for 1ENO/1LNO SL and 200K$<T<$ 260 K for 2ENO/1LNO SL. Interestingly, these are the same temperature ranges, where RTN has been also observed. Further analysis shows that $S_R(f)$ in this temperature range has two components: (a) $1/f$ component, and (b) a Lorentzian term with a corner frequency $f_c$ as
\begin{equation}
\frac{S_R(f)}{R^2}= \frac{A}{f}+\frac{Bf_c}{f^2+f_c^2}
\label{eqn:lorentzian}
\end{equation}
The constants $A$ and $B$ are  the measure of the relative strength of the two terms.  The second term arises from single-frequency fluctuator with a frequency $f_c$.  $f_c$ can be extracted by the fitting (solid line) of the experimental data (symbol)  using Eq. \ref{eqn:lorentzian}, as shown in  Fig. \ref{fig:Fig2}(b) and (c)  for  1ENO/1LNO and 2ENO/1LNO films, respectively for several temperatures.

 The linear relation between  ln$(f_c)$ vs. 1/$T$  (Fig. \ref{fig:Fig2}(d)) demonstrates thermally activated behavior of $f_c$  ($f_c=f_0 e^{-E_a/k_BT}$,  $k_B$ is the Boltzmann's constant)  with an activation energy $E_a\sim$  0.42$\pm$ 0.03 eV for both samples. Similar value of $E_a$ was also observed in   CMR manganite when it undergoes charge order transition~\cite{bid2003low}.  The physical significance of this activated behavior can be visualized  as follows.  For $T<<T_\mathrm{MIT}$, the entire volume is spatially  insulating and the resistance fluctuations are completely random. When the temperature reaches   $T_\mathrm{RTN}$, metallic clusters start to nucleate  in the insulating background. Such metastable metallic phase is separated from the insulating phase by the energy barrier  $E_a$ (upper panel of Fig. \ref{fig:Fig2}(e)) and the competition between these two phases results RTN in the resistance fluctuations. For $T>>T_\mathrm{MIT}$,  the system is again completely in the metallic state (Fig. \ref{fig:Fig2}(e): lower panel) and fluctuations   become random  again. In the subsequent paragraphs, we discuss the results of integrated PSD and second spectrum to strengthen this picture.  Details of the quantitative estimation of noise level can be found in Appendix.

\begin{figure}[b]
	\begin{center}
	\includegraphics[width=0.48\textwidth]{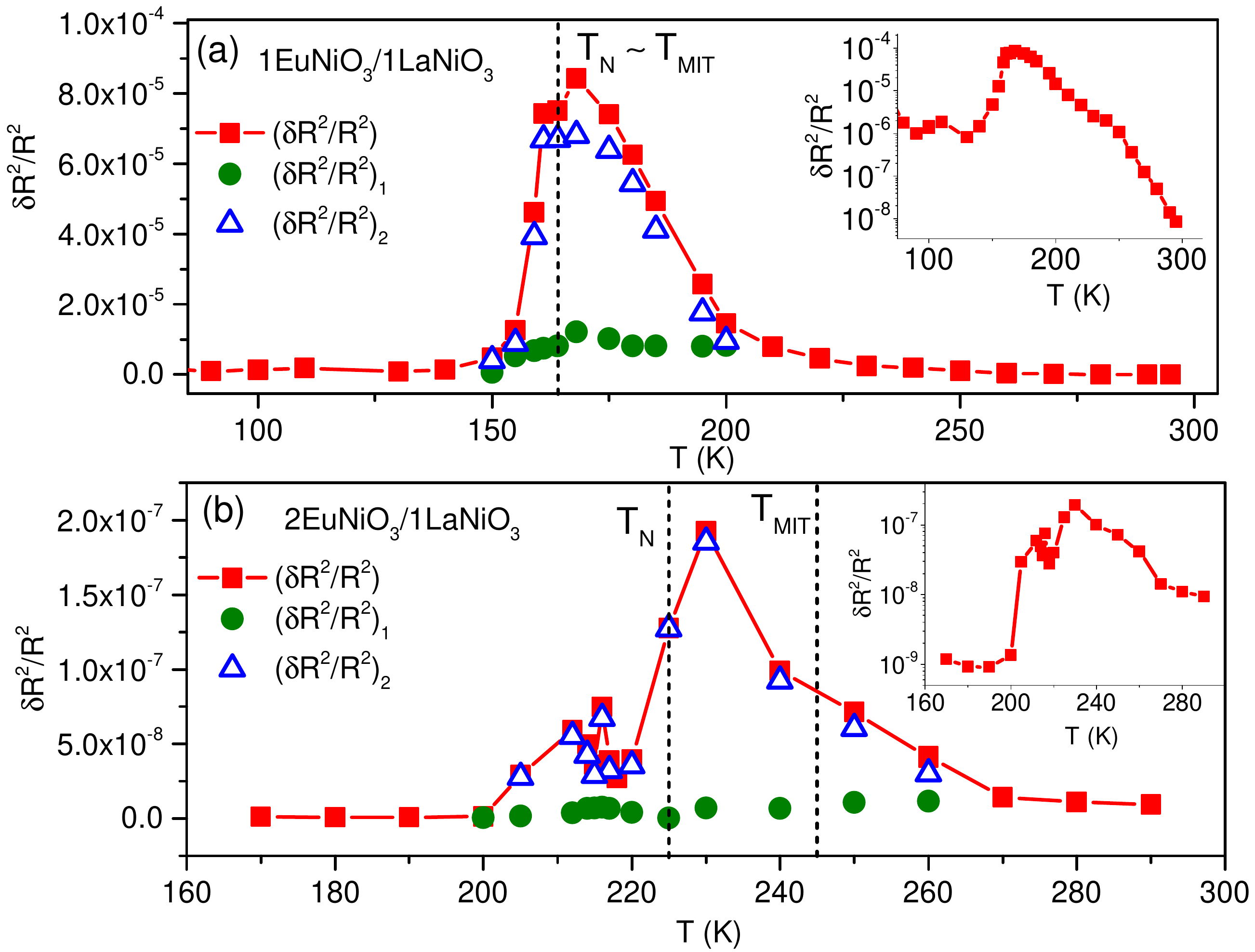}
		\small{\caption{(a) and (b) Temperature dependence total noise $(\frac{\delta R^2}{R^2})$ (red filled squares), $1/f$ noise component $(\frac{\delta R^2}{R^2})_1$ (green filled circles) and Lorentzian component $(\frac{\delta R^2}{R^2})_2$ (blue open triangles) of 1EuNiO$_3$/1LaNiO$_3$ and 2EuNiO$_3$/1LaNiO$_3$ respectively. For details see text. Inset in (a) and (b) shows the semi-log plot of the total noise as a function of temperatures. \label{fig:Fig3}}}
\end{center}
\end{figure}

The PSD of resistance fluctuations has been further integrated over the measurement bandwidth to obtain the relative variance $\frac{\delta R^2}{R^2}$ (noise)  =$\frac{1}{R^2}\int S_R(f)df$.  As can be seen from the inset of Fig. \ref{fig:Fig3} (a) and (b),  the magnitude of noise in the insulating phase of 1ENO/1LNO SL is 10$^3$ times larger than the observed noise in insulating phase of 2ENO/1LNO SL.  The noise value remains almost constant up to  $T \sim 0.85 T_\mathrm{MIT}$ for both samples. While the noise is maximized around   $T_\mathrm{MIT}\sim$ 165 K for 1ENO/1LNO SL, the peak for 2ENO/1LNO SL  appears around 230 K, which is 15 K lower than $T_\mathrm{MIT}\sim$ 245 K.  Interestingly, this 2ENO/1LNO  sample  also shows an additional  noise peak around 210 K, which is again 15 K lower than the $T_N$ $\sim$ 225 K.
 At this moment, the reason for this shift between the transition temperature obtained from resistivity measurement and the  temperature of noise peak remains unclear. It may be related with the resistance fluctuations due to short range charge orderings~\cite{Piamonteze:2005p012104} in insulating phase of this sample.  In spite of  strong difference of the noise magnitude in insulating phases of  1ENO/1LNO and 2ENO/1LNO films,  noise  at 300 K has similar order of magnitude ($\sim 10^{-8}$) in both samples.   In the temperature range $0.85 \lesssim T/T_{MIT} \lesssim$ 1.1,  the total noise $\frac{\delta R^2}{R^2}$ behaves as
$\frac{\delta R^2}{R^2}=\int_{f_{min}}^{f_{max}}\frac{A}{f}df+\int_{f_{min}}^{f_{max}}\frac{Bf_c}{f^2+f_c^2}df
= \bigg(\frac{\delta R^2}{R^2}\bigg)_1+\bigg(\frac{\delta R^2}{R^2}\bigg)_2$  for both samples.
 The temperature dependence of total noise $\frac{\delta R^2}{R^2}$, the contribution of the $1/f$ component $(\frac{\delta R^2}{R^2})_1$ and Lorentzian component $(\frac{\delta R^2}{R^2})_2$ are shown in Fig. \ref{fig:Fig3}(a) and (b).   It is remarkable that noise  in the temperature range $0.85 \lesssim T/T_{MIT} \lesssim  1.1$ predominantly arises from the Lorentzian component  with a  negligible contribution from $1/f$ term.

We note that the noise close to $T_{MIT}$ is 2-4 orders larger than the conventional metal~\cite{hooge1972discussion}, suggesting the microstructural details of the superlattices are different from disorder metallic systems. This large increase in noise close to $T_{MIT}$ can be due to the percolative transition of electrons in an inhomogeneous medium~\cite{PhysRevB.33.2077}.  It has been predicted for such medium from the `random void model' that noise scales as $R^w$ with  $w = 2.1$. We find that  $\frac{\delta R^2}{R^2}$ $\propto$ $R^w$ with $w \sim 2\pm 0.1$ within the temperature range 140 K $\leqslant T \leqslant$ 175 K for 1ENO/1LNO SL and 200 K $\leqslant T \leqslant$ 230 K for 2ENO/1LNO SL (see Appendix: Fig. \ref{fig:Fig6}). Such classical percolation picture has  been reported in other oxides as well when they undergo normal to superconducting phase transition~\cite{kiss1994noise,daptary2016correlated}. It can be seen from Fig. \ref{fig:Fig3} that the magnitude of noise of 1ENO/1LNO is three orders larger than that of 2ENO/1LNO. Large increase in noise have been seen in other oxide undergoing long range charge-ordering transition~\cite{bid2003low}. We speculate that because of the absence of long range CO in 2ENO/1LNO SL,  noise magnitude is smaller than 1ENO/1LNO SL.

\begin{figure}[t!]
\begin{center}
\includegraphics[width=0.48\textwidth]{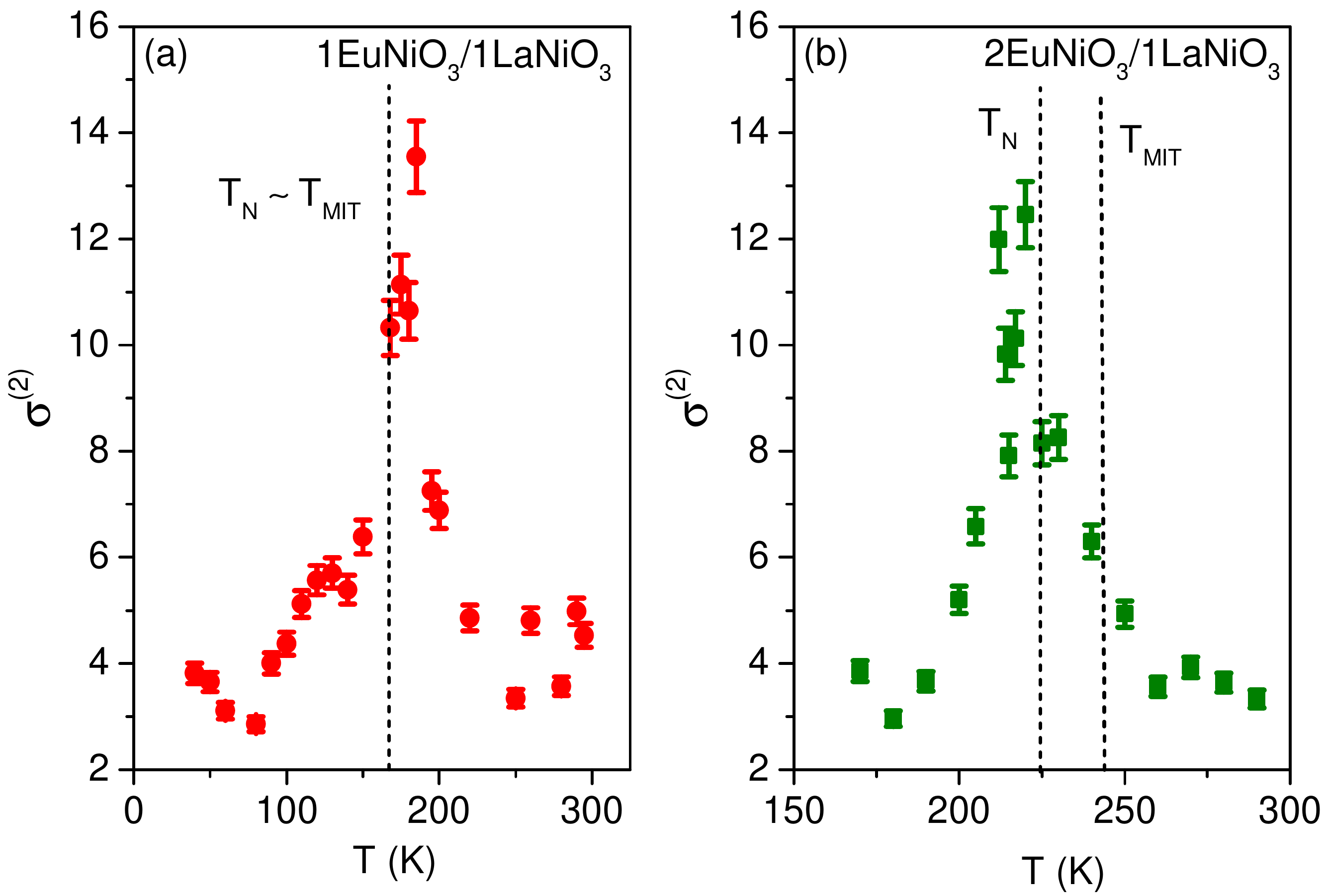}
\small{\caption{(a) and (b) Plot of normalized second spectrum $\sigma^{(2)}$ as a function of temperature of 1EuNiO$_3$/1LaNiO$_3$ and 2EuNiO$_3$/1LaNiO$_3$ respectively.  \label{fig:Fig4}}}
\end{center}
\end{figure}

 While the origin of noise in metallic phase can be understood from the Dutta-Horn model ~\cite{dutta1981low,daptary2017effect}, such
 defect scattering based mechanism fails to explain the peculiar behavior of noise in the temperature range $0.85 \lesssim T/T_{MIT} \lesssim  1.1$ region. To gain better understanding of the origin of such excess noise, we have investigated higher order statistics of resistance fluctuations, which have been used to study the presence of long-range correlations undergoing magnetic, spin-glass \cite{daptary2014probing}, superconducting transition \cite{daptary2016correlated}. To calculate the higher order statistics of resistance fluctuations, we have computed the second spectrum. The second spectrum is a four-point correlation function of the resistance fluctuations over a chosen octave ($f_l, f_h$) and is   defined as
$S_R^{f_1}(f_2)=\int_0^\infty \langle\delta R^2(t)\rangle\langle\delta R^2(t+\tau)\rangle cos(2\pi f_2\tau)d\tau$
where $f_1$ is the centre frequency of a chosen octave and $f_2$ is the spectral frequency. Physically, $S_R^{f_1}(f_2)$ represents the \textquoteleft spectral wandering\textquoteright or fluctuations in the PSD with time in the chosen frequency octave~\cite{seidler1996non}.  To avoid   artifacts in the actual signal from the Gaussian background noise, we have calculated the second
spectrum over the frequency octave 0.09375-0.1875 Hz, where the sample noise is significantly higher than the background noise. We show plot of the variation of second spectrum $S_R^{f_1}(f_2)$ with of frequency at different $T$ for both superlattices in Appendix (Fig.~\ref{fig:Fig7}). A convenient way of representing the second spectrum is through the normalized form $\sigma^{(2)}$ defined as
$\sigma^{(2)}=\int_0^{f_h-f_l}S_R^{f_1}(f_2)df_2/[\int_{f_l}^{f_h}S_R(f)df]^2$.
For Gaussian fluctuations, $\sigma^{(2)}$ = 3 and any deviation from this value would imply the presence of NGC in the fluctuation spectrum~\cite{seidler1996non}. As expected,  $\sigma^{(2)} \sim 3$ in metallic phase of both samples  (Fig. \ref{fig:Fig4}(a) and (b)).   $\sigma^{(2)}$ starts to deviate from 3 for both samples around $T_\mathrm{MIT}$, implying that the observation of excess noise is intimately connected to the electronic phase separation. In case of 1ENO/1LNO SL, $\sigma^{(2)}$ shows a peak near $T\sim T_\mathrm{MIT}=T_N$. On the contrary, $\sigma^{(2)}$ becomes maximum near $T\sim T_N$ for 2ENO/1LNO SL. This surprising observation is likely to be   related to the multi-band nature of these materials.  The Fermi surface of the paramagnetic metallic phase consists of large hole pockets with small electron pockets~\cite{Eguchi:2009p115122}. As inferred from the Hall effect measurements by Ojha et al.~\cite{Ojha2019}, the metal insulator transition results in a partially gapped Fermi surface  and the hole Fermi surface vanishes around $T_N$ by the nesting driven  paramagnetic to $E^\prime$-antiferromagnetic transition.

The length scale associated with the  electronic phase separation can be estimated  if we consider that  the activation energy ($E_a$)  corresponds to the pure elastic energy generated due to the volume difference between metallic and insulating phase~\cite{bid2003low}. The bulk modulus of EuNiO$_3$ and LaNiO$_3$ is approximately 320 GPa and 380 GPa, respectively~\cite{Zhou:2004p081102}.  The out-of-plane lattice constant of the SLs shows around 0.2\% expansion across the MIT~\cite{Middey:2018p156801},  yielding  an energy density ($E_v$)  $\sim$ 160-190 kJ/m$^3$ associated with the transformation. By assuming that the metallic nucleating regions   are spherical with a diameter $L_m$,   $E_a\sim$ 0.42 eV corresponds to $L_m\sim$ 7.0-7.4 nm.  We note that conductive-atomic force microscopy study with a spatial resolution of $\sim$ 100 nm has found  nucleation of metallic domains with size $\sim$100-300 nm in a NdNiO$_3$ thin film~\cite{preziosi:2018p2226}.  Our  results  emphasizes that nucleation of such metallic phase happens   at much shorter length scale.

 Earlier X-ray absorption spectroscopy experiment demonstrated the presence of short-range charge ordering even in metallic phase of all $RE$NiO$_3$~\cite{Piamonteze:2005p012104}.
Our present noise measurements  emphasize  similar characteristics for both samples, such as random resistance fluctuations, 1/$f$ noise for $T>T_\mathrm{MIT}$ and RTN in resistance fluctuations, non-1/$f$  and non-Gaussian characterization of noise for  $T>0.85T_\mathrm{MIT}$. Further,  the very similar length scale associated with  nucleation of metallic clusters in the insulating phase in  both samples also suggests that the samples have  similar electronic and magnetic properties in nanoscale. However,  the much smaller noise magnitude of 2ENO/1LNO around $T_\mathrm{MIT}$ compared to 1ENO/1LNO SL,  and  additional noise peak near $T_N$  infer that  the details of the electrical transport process  depend on the presence/absence of long-range charge  and magnetic orderings. Complimentary microscopy experiments with sub-nm resolution should   help to clarify the  details of the development of long-range charge ordered phase from a  short-range charge ordered phase and  the  magnetic nature of phase separated phases around the transition temperatures.

\section{Conclusion}
To summarize, we have observed the presence of large excess noise around the metal-insulator   and magnetic transitions in EuNiO$_3$/LaNiO$_3$ thin films. The appearance of RTN, causing non-1/$f$ noise below $T_\mathrm{MIT}$ implies that the electronic phase separation is   responsible for the excess noise. This is further corroborated by the observation of a large non-Gaussian noise in the insulating phase. Noise in the metallic phase shows 1/$f$ behavior with the Gaussian statistics of the resistance fluctuations.   Observation of the maxima of $\sigma^{(2)}$ near the $T_N$ for 2ENO/1LNO SL is likely to be connected to the Fermi surface nesting driven origin of  $E^\prime$-type antiferromagnetic ordering.  Our experiments highlight the importance of resistance fluctuations study near the phase transition which can be applied to understand  transition between  two  electronic phases of any system.

\section{Acknowledgements}

SM acknowledges IISc start up grant and DST Nanomission grant  no. DST/NM/NS/2018/246 for  financial support. AB thanks SERB, DST for financial support. J.C. is supported by the Gordon and Betty Moore Foundation EPiQS Initiative through Grant No. GBMF4534.

\section*{Appendix}

To compare the noise level of EuNiO$_3$/LaNiO$_3$ with other nickelates, we have calculated the Hooge parameter $\gamma_H$ \cite{hooge19691}, defined as $\gamma_H=\frac{N\times f \times S_R(f)}{R^2}$ ($N$, total number of charge carriers  has been evaluated from Hall effect measurement). The value of $\gamma_H$ for different nickelates have been tabulated in table \ref{tab:oxides}. The large value of $\gamma_H$ suggests that the origin of noise of these $RE$NiO$_3$ are different from the scattering mechanism of electron with lattice phonon mode predicted by Hooge for metals and semiconductors \cite{hooge1972discussion}. A possible explanation of large increase in noise is the classical percolation of electrons in inhomogeneous medium~\cite{PhysRevB.33.2077}.
 
\begin{table}[h]
\centering
	\small{\caption{Value of Hooge parameter for different nickelates at 300 K. \label{tab:oxides}}}
\begin{tabular}{| l | c | c | c | c | c | c | c | c | }
\hline
System &Hooge parameter at 300 K\\ \hline
		1EuNiO$_3$/1LaNiO$_3$ &10$^4$ \\ \hline
		2EuNiO$_3$/1LaNiO$_3$ &10$^4$ \\ \hline
		LaNiO$_{3-\delta}$~\cite{ghosh1997dependence} &10$^3$ \\ \hline
		SmNiO$_3$~\cite{sahoo2014conductivity} &$5 \times 10^3$ \\ \hline
		NdNiO$_3$~\cite{Bisht:2017p115147,alsaqqa2017phase} &$5 \times 10^6$\\ \hline
		\end{tabular}

	
\end{table}

\begin{figure}[b!]
\begin{center}
\includegraphics[width=0.4\textwidth]{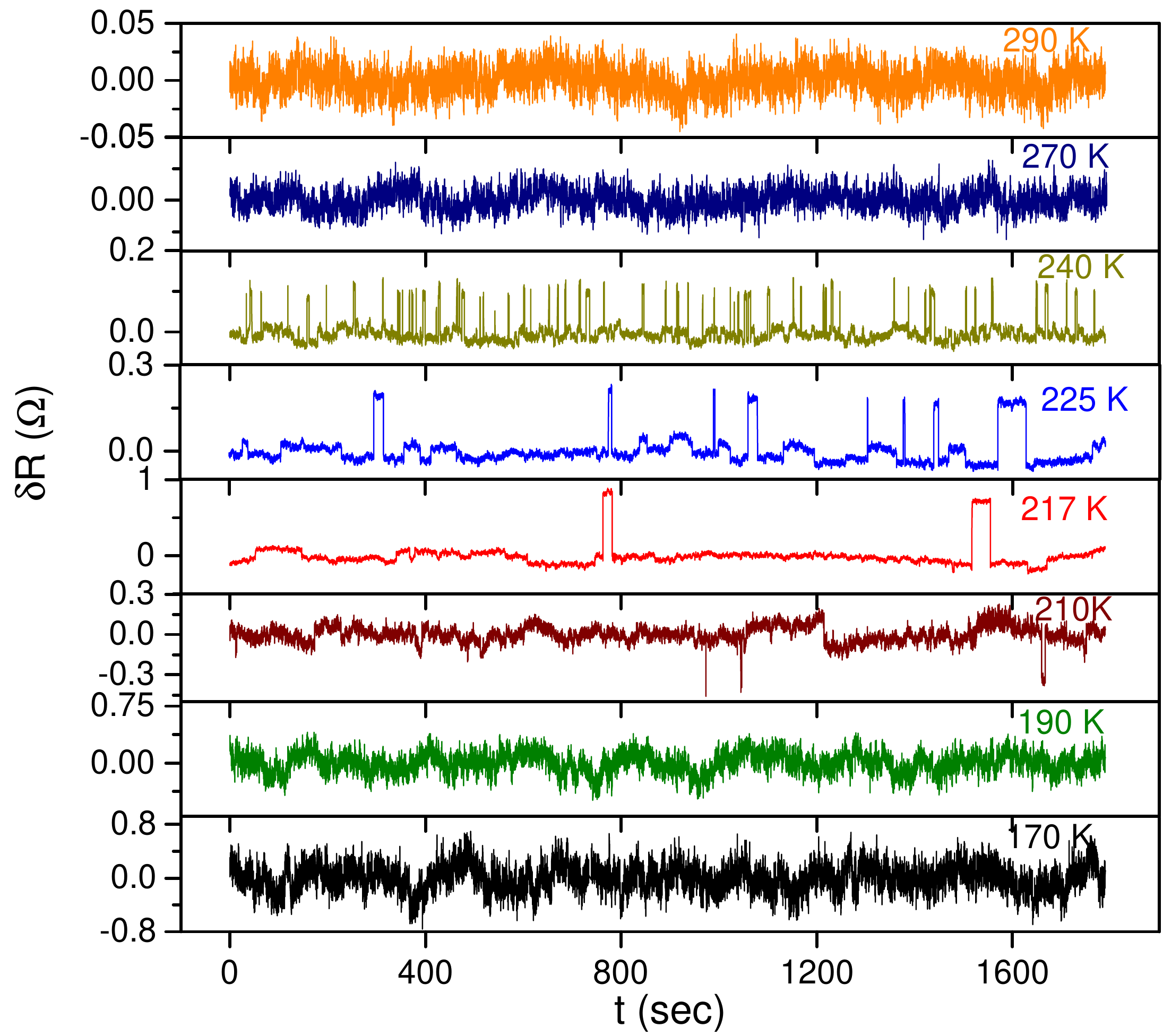}
\small{\caption{ Time series of resistance fluctuations at few representative values of $T$ of 2EuNiO$_3$/1LaNiO$_3$.  \label{fig:Fig5}}}
\end{center}
\end{figure}

\begin{figure}
\begin{center}
\includegraphics[width=0.4\textwidth]{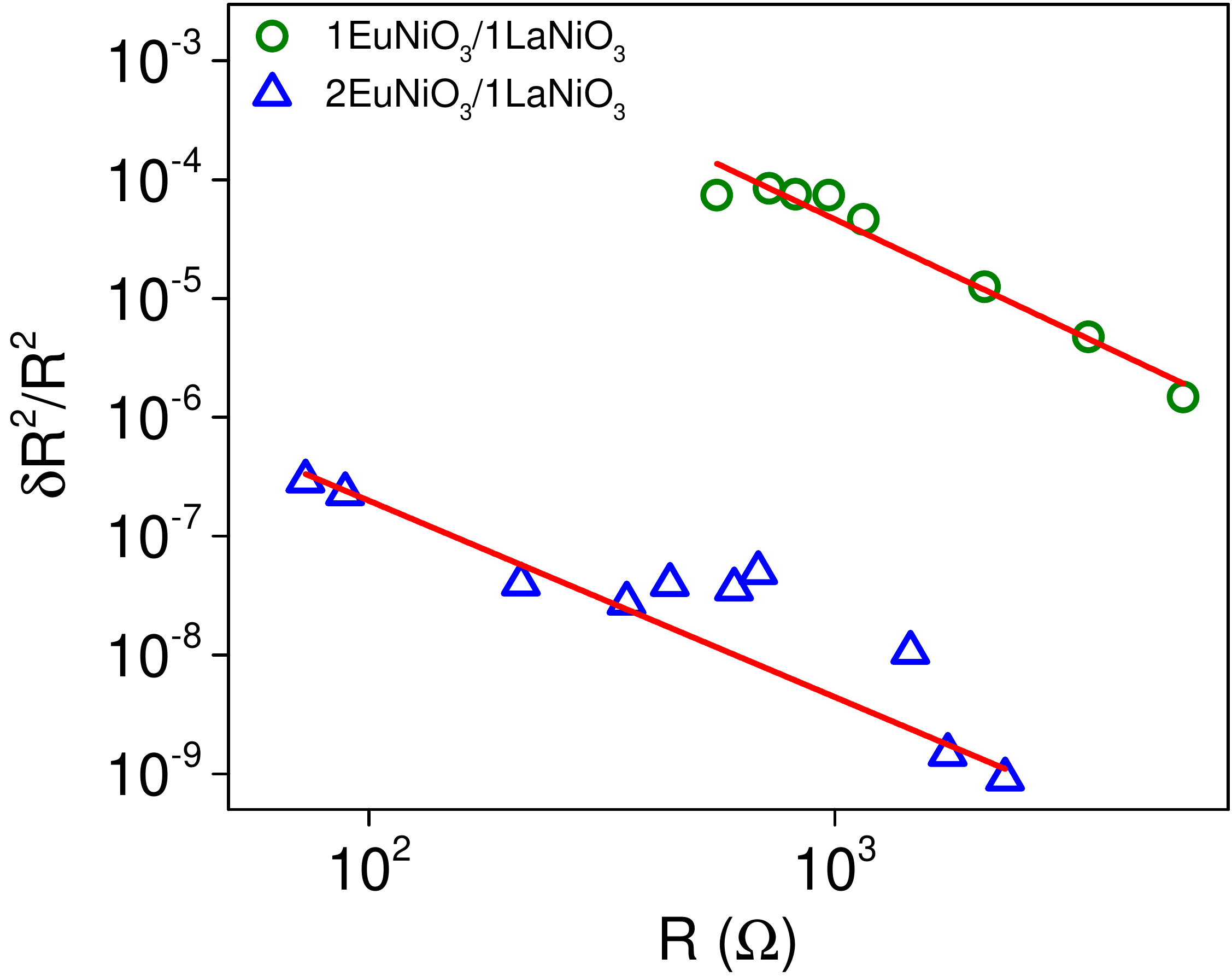}
\small{\caption{ Log-log plot of $\frac{\delta R^2}{R^2}$ as a function of resistance of 1EuNiO$_3$/1LaNiO$_3$ and 2EuNiO$_3$/1LaNiO$_3$ respectively. The red line represents the linear fit of the data of slope $2\pm 0.1$. For details see text. \label{fig:Fig6}}}
\end{center}
\end{figure}

\begin{figure}
\begin{center}
\includegraphics[width=0.48\textwidth]{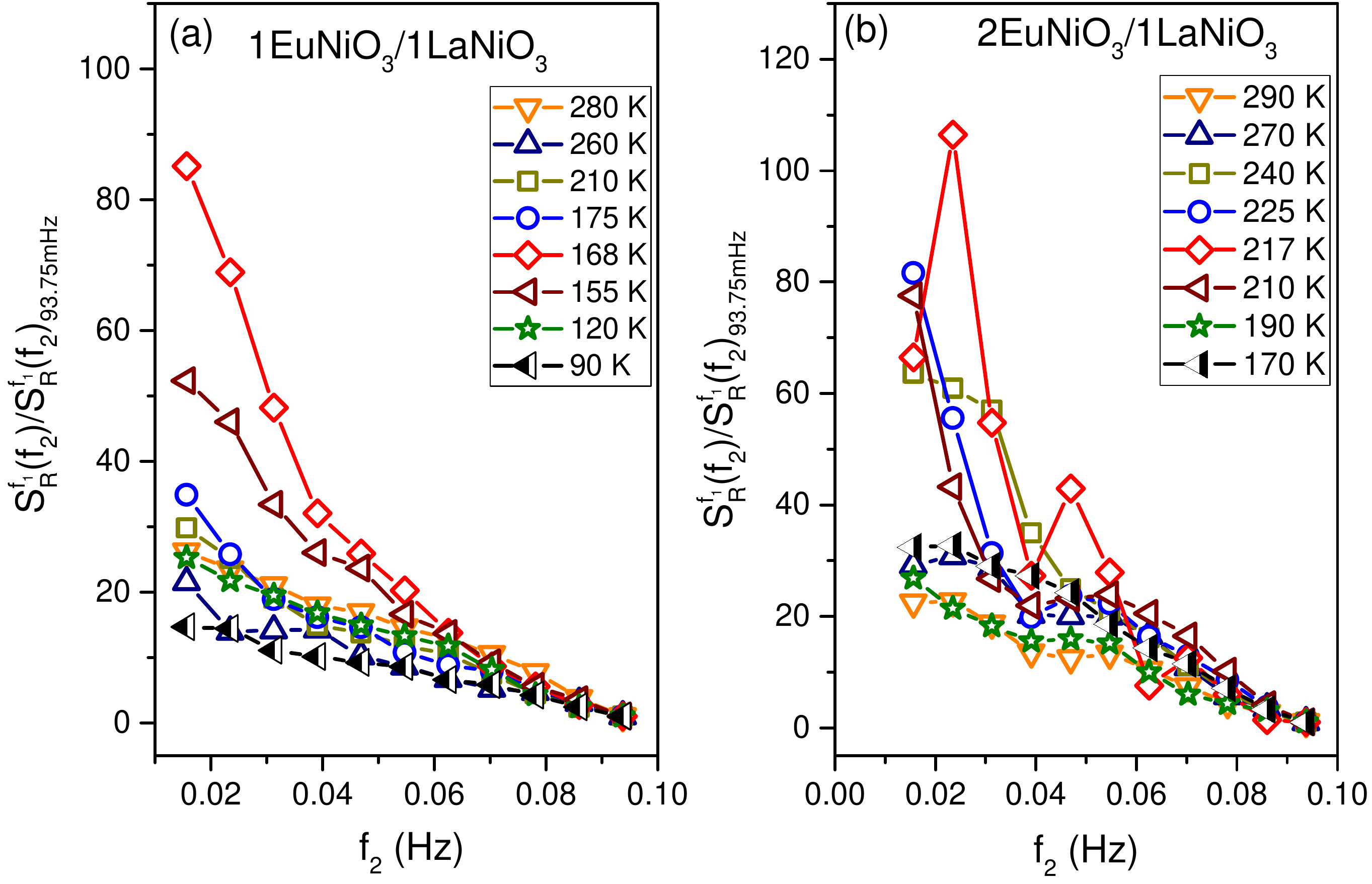}
\small{\caption{(a) and (b) Plot of second spectrum $S_R^{f_1}(f_2)$ as a function of frequency at different $T$ for 1EuNiO$_3$/1LaNiO$_3$ and 2EuNiO$_3$/1LaNiO$_3$, respectively. Note that the data have been scaled at 93.75 mHz to see the nature at different temperatures. \label{fig:Fig7}}}
\end{center}
\end{figure}

\newpage

\begin{thebibliography}{53}%
\makeatletter
\providecommand \@ifxundefined [1]{%
 \@ifx{#1\undefined}
}%
\providecommand \@ifnum [1]{%
 \ifnum #1\expandafter \@firstoftwo
 \else \expandafter \@secondoftwo
 \fi
}%
\providecommand \@ifx [1]{%
 \ifx #1\expandafter \@firstoftwo
 \else \expandafter \@secondoftwo
 \fi
}%
\providecommand \natexlab [1]{#1}%
\providecommand \enquote  [1]{``#1''}%
\providecommand \bibnamefont  [1]{#1}%
\providecommand \bibfnamefont [1]{#1}%
\providecommand \citenamefont [1]{#1}%
\providecommand \href@noop [0]{\@secondoftwo}%
\providecommand \href [0]{\begingroup \@sanitize@url \@href}%
\providecommand \@href[1]{\@@startlink{#1}\@@href}%
\providecommand \@@href[1]{\endgroup#1\@@endlink}%
\providecommand \@sanitize@url [0]{\catcode `\\12\catcode `\$12\catcode
  `\&12\catcode `\#12\catcode `\^12\catcode `\_12\catcode `\%12\relax}%
\providecommand \@@startlink[1]{}%
\providecommand \@@endlink[0]{}%
\providecommand \url  [0]{\begingroup\@sanitize@url \@url }%
\providecommand \@url [1]{\endgroup\@href {#1}{\urlprefix }}%
\providecommand \urlprefix  [0]{URL }%
\providecommand \Eprint [0]{\href }%
\providecommand \doibase [0]{http://dx.doi.org/}%
\providecommand \selectlanguage [0]{\@gobble}%
\providecommand \bibinfo  [0]{\@secondoftwo}%
\providecommand \bibfield  [0]{\@secondoftwo}%
\providecommand \translation [1]{[#1]}%
\providecommand \BibitemOpen [0]{}%
\providecommand \bibitemStop [0]{}%
\providecommand \bibitemNoStop [0]{.\EOS\space}%
\providecommand \EOS [0]{\spacefactor3000\relax}%
\providecommand \BibitemShut  [1]{\csname bibitem#1\endcsname}%
\let\auto@bib@innerbib\@empty
\bibitem [{\citenamefont {Imada}\ \emph {et~al.}(1998)\citenamefont {Imada},
  \citenamefont {Fujimori},\ and\ \citenamefont {Tokura}}]{Imada:1998p1039}%
  \BibitemOpen
  \bibfield  {author} {\bibinfo {author} {\bibfnamefont {M.}~\bibnamefont
  {Imada}}, \bibinfo {author} {\bibfnamefont {A.}~\bibnamefont {Fujimori}}, \
  and\ \bibinfo {author} {\bibfnamefont {Y.}~\bibnamefont {Tokura}},\ }\href
  {\doibase 10.1103/RevModPhys.70.1039} {\bibfield  {journal} {\bibinfo
  {journal} {Rev. Mod. Phys.}\ }\textbf {\bibinfo {volume} {70}},\ \bibinfo
  {pages} {1039} (\bibinfo {year} {1998})}\BibitemShut {NoStop}%
\bibitem [{\citenamefont {Tokura}(2006)}]{Tokura:2006p797}%
  \BibitemOpen
  \bibfield  {author} {\bibinfo {author} {\bibfnamefont {Y.}~\bibnamefont
  {Tokura}},\ }\href {\doibase 10.1088/0034-4885/69/3/r06} {\bibfield
  {journal} {\bibinfo  {journal} {Reports on Progress in Physics}\ }\textbf
  {\bibinfo {volume} {69}},\ \bibinfo {pages} {797} (\bibinfo {year}
  {2006})}\BibitemShut {NoStop}%
\bibitem [{\citenamefont {Middey}\ \emph {et~al.}(2016)\citenamefont {Middey},
  \citenamefont {Chakhalian}, \citenamefont {Mahadevan}, \citenamefont
  {Freeland}, \citenamefont {Millis},\ and\ \citenamefont
  {Sarma}}]{Middey:2016p305}%
  \BibitemOpen
  \bibfield  {author} {\bibinfo {author} {\bibfnamefont {S.}~\bibnamefont
  {Middey}}, \bibinfo {author} {\bibfnamefont {J.}~\bibnamefont {Chakhalian}},
  \bibinfo {author} {\bibfnamefont {P.}~\bibnamefont {Mahadevan}}, \bibinfo
  {author} {\bibfnamefont {J.~W.}\ \bibnamefont {Freeland}}, \bibinfo {author}
  {\bibfnamefont {A.~J.}\ \bibnamefont {Millis}}, \ and\ \bibinfo {author}
  {\bibfnamefont {D.~D.}\ \bibnamefont {Sarma}},\ }\href {\doibase
  10.1146/annurev-matsci-070115-032057} {\bibfield  {journal} {\bibinfo
  {journal} {Annual Review of Materials Research}\ }\textbf {\bibinfo {volume}
  {46}},\ \bibinfo {pages} {305} (\bibinfo {year} {2016})}\BibitemShut
  {NoStop}%
\bibitem [{\citenamefont {Catalano}\ \emph {et~al.}(2018)\citenamefont
  {Catalano}, \citenamefont {Gibert}, \citenamefont {Fowlie}, \citenamefont
  {{\'I}{\~n}iguez}, \citenamefont {Triscone},\ and\ \citenamefont
  {Kreisel}}]{Catalano:2018p046501}%
  \BibitemOpen
  \bibfield  {author} {\bibinfo {author} {\bibfnamefont {S.}~\bibnamefont
  {Catalano}}, \bibinfo {author} {\bibfnamefont {M.}~\bibnamefont {Gibert}},
  \bibinfo {author} {\bibfnamefont {J.}~\bibnamefont {Fowlie}}, \bibinfo
  {author} {\bibfnamefont {J.}~\bibnamefont {{\'I}{\~n}iguez}}, \bibinfo
  {author} {\bibfnamefont {J.-M.}\ \bibnamefont {Triscone}}, \ and\ \bibinfo
  {author} {\bibfnamefont {J.}~\bibnamefont {Kreisel}},\ }\href {\doibase
  10.1088/1361-6633/aaa37a} {\bibfield  {journal} {\bibinfo  {journal} {Reports
  on Progress in Physics}\ }\textbf {\bibinfo {volume} {81}},\ \bibinfo {pages}
  {046501} (\bibinfo {year} {2018})}\BibitemShut {NoStop}%
\bibitem [{\citenamefont {Medarde}(1997)}]{Medarde:1997p1679}%
  \BibitemOpen
  \bibfield  {author} {\bibinfo {author} {\bibfnamefont {M.~L.}\ \bibnamefont
  {Medarde}},\ }\href@noop {} {\bibfield  {journal} {\bibinfo  {journal}
  {Journal of Physics: Condensed Matter}\ }\textbf {\bibinfo {volume} {9}},\
  \bibinfo {pages} {1679} (\bibinfo {year} {1997})}\BibitemShut {NoStop}%
\bibitem [{\citenamefont {Catalan}(2008)}]{Catalan:2008p729}%
  \BibitemOpen
  \bibfield  {author} {\bibinfo {author} {\bibfnamefont {G.}~\bibnamefont
  {Catalan}},\ }\href {\doibase 10.1080/01411590801992463} {\bibfield
  {journal} {\bibinfo  {journal} {Phase Transitions}\ }\textbf {\bibinfo
  {volume} {81}},\ \bibinfo {pages} {729} (\bibinfo {year} {2008})}\BibitemShut
  {NoStop}%
\bibitem [{\citenamefont {Mercy}\ \emph {et~al.}(2017)\citenamefont {Mercy},
  \citenamefont {Bieder}, \citenamefont {{\'I}{\~n}iguez},\ and\ \citenamefont
  {Ghosez}}]{Mercy:2017p1667}%
  \BibitemOpen
  \bibfield  {author} {\bibinfo {author} {\bibfnamefont {A.}~\bibnamefont
  {Mercy}}, \bibinfo {author} {\bibfnamefont {J.}~\bibnamefont {Bieder}},
  \bibinfo {author} {\bibfnamefont {J.}~\bibnamefont {{\'I}{\~n}iguez}}, \ and\
  \bibinfo {author} {\bibfnamefont {P.}~\bibnamefont {Ghosez}},\ }\href
  {\doibase 10.1038/s41467-017-01811-x} {\bibfield  {journal} {\bibinfo
  {journal} {Nature Communications}\ }\textbf {\bibinfo {volume} {8}},\
  \bibinfo {pages} {1667} (\bibinfo {year} {2017})}\BibitemShut {NoStop}%
\bibitem [{\citenamefont {Stewart}\ \emph {et~al.}(2011)\citenamefont
  {Stewart}, \citenamefont {Liu}, \citenamefont {Kareev}, \citenamefont
  {Chakhalian},\ and\ \citenamefont {Basov}}]{Stewart:2011p176401}%
  \BibitemOpen
  \bibfield  {author} {\bibinfo {author} {\bibfnamefont {M.~K.}\ \bibnamefont
  {Stewart}}, \bibinfo {author} {\bibfnamefont {J.}~\bibnamefont {Liu}},
  \bibinfo {author} {\bibfnamefont {M.}~\bibnamefont {Kareev}}, \bibinfo
  {author} {\bibfnamefont {J.}~\bibnamefont {Chakhalian}}, \ and\ \bibinfo
  {author} {\bibfnamefont {D.~N.}\ \bibnamefont {Basov}},\ }\href {\doibase
  10.1103/PhysRevLett.107.176401} {\bibfield  {journal} {\bibinfo  {journal}
  {Phys. Rev. Lett.}\ }\textbf {\bibinfo {volume} {107}},\ \bibinfo {pages}
  {176401} (\bibinfo {year} {2011})}\BibitemShut {NoStop}%
\bibitem [{\citenamefont {Staub}\ \emph {et~al.}(2002)\citenamefont {Staub},
  \citenamefont {Meijer}, \citenamefont {Fauth}, \citenamefont {Allenspach},
  \citenamefont {Bednorz}, \citenamefont {Karpinski}, \citenamefont {Kazakov},
  \citenamefont {Paolasini},\ and\ \citenamefont
  {d'Acapito}}]{Staub:2002p126402}%
  \BibitemOpen
  \bibfield  {author} {\bibinfo {author} {\bibfnamefont {U.}~\bibnamefont
  {Staub}}, \bibinfo {author} {\bibfnamefont {G.~I.}\ \bibnamefont {Meijer}},
  \bibinfo {author} {\bibfnamefont {F.}~\bibnamefont {Fauth}}, \bibinfo
  {author} {\bibfnamefont {R.}~\bibnamefont {Allenspach}}, \bibinfo {author}
  {\bibfnamefont {J.~G.}\ \bibnamefont {Bednorz}}, \bibinfo {author}
  {\bibfnamefont {J.}~\bibnamefont {Karpinski}}, \bibinfo {author}
  {\bibfnamefont {S.~M.}\ \bibnamefont {Kazakov}}, \bibinfo {author}
  {\bibfnamefont {L.}~\bibnamefont {Paolasini}}, \ and\ \bibinfo {author}
  {\bibfnamefont {F.}~\bibnamefont {d'Acapito}},\ }\href {\doibase
  10.1103/PhysRevLett.88.126402} {\bibfield  {journal} {\bibinfo  {journal}
  {Phys. Rev. Lett.}\ }\textbf {\bibinfo {volume} {88}},\ \bibinfo {pages}
  {126402} (\bibinfo {year} {2002})}\BibitemShut {NoStop}%
\bibitem [{\citenamefont {Mazin}\ \emph {et~al.}(2007)\citenamefont {Mazin},
  \citenamefont {Khomskii}, \citenamefont {Lengsdorf}, \citenamefont {Alonso},
  \citenamefont {Marshall}, \citenamefont {Ibberson}, \citenamefont
  {Podlesnyak}, \citenamefont {Mart\'{\i}nez-Lope},\ and\ \citenamefont
  {Abd-Elmeguid}}]{Mazin:2007p176406}%
  \BibitemOpen
  \bibfield  {author} {\bibinfo {author} {\bibfnamefont {I.~I.}\ \bibnamefont
  {Mazin}}, \bibinfo {author} {\bibfnamefont {D.~I.}\ \bibnamefont {Khomskii}},
  \bibinfo {author} {\bibfnamefont {R.}~\bibnamefont {Lengsdorf}}, \bibinfo
  {author} {\bibfnamefont {J.~A.}\ \bibnamefont {Alonso}}, \bibinfo {author}
  {\bibfnamefont {W.~G.}\ \bibnamefont {Marshall}}, \bibinfo {author}
  {\bibfnamefont {R.~M.}\ \bibnamefont {Ibberson}}, \bibinfo {author}
  {\bibfnamefont {A.}~\bibnamefont {Podlesnyak}}, \bibinfo {author}
  {\bibfnamefont {M.~J.}\ \bibnamefont {Mart\'{\i}nez-Lope}}, \ and\ \bibinfo
  {author} {\bibfnamefont {M.~M.}\ \bibnamefont {Abd-Elmeguid}},\ }\href
  {\doibase 10.1103/PhysRevLett.98.176406} {\bibfield  {journal} {\bibinfo
  {journal} {Phys. Rev. Lett.}\ }\textbf {\bibinfo {volume} {98}},\ \bibinfo
  {pages} {176406} (\bibinfo {year} {2007})}\BibitemShut {NoStop}%
\bibitem [{\citenamefont {Barman}\ \emph {et~al.}(1994)\citenamefont {Barman},
  \citenamefont {Chainani},\ and\ \citenamefont {Sarma}}]{Barman:1994p8475}%
  \BibitemOpen
  \bibfield  {author} {\bibinfo {author} {\bibfnamefont {S.~R.}\ \bibnamefont
  {Barman}}, \bibinfo {author} {\bibfnamefont {A.}~\bibnamefont {Chainani}}, \
  and\ \bibinfo {author} {\bibfnamefont {D.~D.}\ \bibnamefont {Sarma}},\ }\href
  {\doibase 10.1103/PhysRevB.49.8475} {\bibfield  {journal} {\bibinfo
  {journal} {Phys. Rev. B}\ }\textbf {\bibinfo {volume} {49}},\ \bibinfo
  {pages} {8475} (\bibinfo {year} {1994})}\BibitemShut {NoStop}%
\bibitem [{\citenamefont {Mizokawa}\ \emph {et~al.}(2000)\citenamefont
  {Mizokawa}, \citenamefont {Khomskii},\ and\ \citenamefont
  {Sawatzky}}]{Mizokawa2000p11263}%
  \BibitemOpen
  \bibfield  {author} {\bibinfo {author} {\bibfnamefont {T.}~\bibnamefont
  {Mizokawa}}, \bibinfo {author} {\bibfnamefont {D.~I.}\ \bibnamefont
  {Khomskii}}, \ and\ \bibinfo {author} {\bibfnamefont {G.~A.}\ \bibnamefont
  {Sawatzky}},\ }\href {\doibase 10.1103/PhysRevB.61.11263} {\bibfield
  {journal} {\bibinfo  {journal} {Phys. Rev. B}\ }\textbf {\bibinfo {volume}
  {61}},\ \bibinfo {pages} {11263} (\bibinfo {year} {2000})}\BibitemShut
  {NoStop}%
\bibitem [{\citenamefont {Park}\ \emph {et~al.}(2012)\citenamefont {Park},
  \citenamefont {Millis},\ and\ \citenamefont {Marianetti}}]{Park:2012p156402}%
  \BibitemOpen
  \bibfield  {author} {\bibinfo {author} {\bibfnamefont {H.}~\bibnamefont
  {Park}}, \bibinfo {author} {\bibfnamefont {A.~J.}\ \bibnamefont {Millis}}, \
  and\ \bibinfo {author} {\bibfnamefont {C.~A.}\ \bibnamefont {Marianetti}},\
  }\href {\doibase 10.1103/PhysRevLett.109.156402} {\bibfield  {journal}
  {\bibinfo  {journal} {Phys. Rev. Lett.}\ }\textbf {\bibinfo {volume} {109}},\
  \bibinfo {pages} {156402} (\bibinfo {year} {2012})}\BibitemShut {NoStop}%
\bibitem [{\citenamefont {Johnston}\ \emph {et~al.}(2014)\citenamefont
  {Johnston}, \citenamefont {Mukherjee}, \citenamefont {Elfimov}, \citenamefont
  {Berciu},\ and\ \citenamefont {Sawatzky}}]{johnston:2014p106404}%
  \BibitemOpen
  \bibfield  {author} {\bibinfo {author} {\bibfnamefont {S.}~\bibnamefont
  {Johnston}}, \bibinfo {author} {\bibfnamefont {A.}~\bibnamefont {Mukherjee}},
  \bibinfo {author} {\bibfnamefont {I.}~\bibnamefont {Elfimov}}, \bibinfo
  {author} {\bibfnamefont {M.}~\bibnamefont {Berciu}}, \ and\ \bibinfo {author}
  {\bibfnamefont {G.~A.}\ \bibnamefont {Sawatzky}},\ }\href {\doibase
  10.1103/PhysRevLett.112.106404} {\bibfield  {journal} {\bibinfo  {journal}
  {Phys. Rev. Lett.}\ }\textbf {\bibinfo {volume} {112}},\ \bibinfo {pages}
  {106404} (\bibinfo {year} {2014})}\BibitemShut {NoStop}%
\bibitem [{\citenamefont {Subedi}\ \emph {et~al.}(2015)\citenamefont {Subedi},
  \citenamefont {Peil},\ and\ \citenamefont {Georges}}]{Subedi:2015p075128}%
  \BibitemOpen
  \bibfield  {author} {\bibinfo {author} {\bibfnamefont {A.}~\bibnamefont
  {Subedi}}, \bibinfo {author} {\bibfnamefont {O.~E.}\ \bibnamefont {Peil}}, \
  and\ \bibinfo {author} {\bibfnamefont {A.}~\bibnamefont {Georges}},\ }\href
  {\doibase 10.1103/PhysRevB.91.075128} {\bibfield  {journal} {\bibinfo
  {journal} {Phys. Rev. B}\ }\textbf {\bibinfo {volume} {91}},\ \bibinfo
  {pages} {075128} (\bibinfo {year} {2015})}\BibitemShut {NoStop}%
\bibitem [{\citenamefont {Bisogni}\ \emph {et~al.}(2016)\citenamefont
  {Bisogni}, \citenamefont {Catalano}, \citenamefont {Green}, \citenamefont
  {Gibert}, \citenamefont {Scherwitzl}, \citenamefont {Huang}, \citenamefont
  {Strocov}, \citenamefont {Zubko}, \citenamefont {Balandeh}, \citenamefont
  {Triscone},\ and\ \citenamefont {et~al.}}]{Bisogni:2016p13017}%
  \BibitemOpen
  \bibfield  {author} {\bibinfo {author} {\bibfnamefont {V.}~\bibnamefont
  {Bisogni}}, \bibinfo {author} {\bibfnamefont {S.}~\bibnamefont {Catalano}},
  \bibinfo {author} {\bibfnamefont {R.~J.}\ \bibnamefont {Green}}, \bibinfo
  {author} {\bibfnamefont {M.}~\bibnamefont {Gibert}}, \bibinfo {author}
  {\bibfnamefont {R.}~\bibnamefont {Scherwitzl}}, \bibinfo {author}
  {\bibfnamefont {Y.}~\bibnamefont {Huang}}, \bibinfo {author} {\bibfnamefont
  {V.~N.}\ \bibnamefont {Strocov}}, \bibinfo {author} {\bibfnamefont
  {P.}~\bibnamefont {Zubko}}, \bibinfo {author} {\bibfnamefont
  {S.}~\bibnamefont {Balandeh}}, \bibinfo {author} {\bibfnamefont {J.-M.}\
  \bibnamefont {Triscone}}, \ and\ \bibinfo {author} {\bibnamefont {et~al.}},\
  }\href {\doibase 10.1038/ncomms13017} {\bibfield  {journal} {\bibinfo
  {journal} {Nature Communications}\ }\textbf {\bibinfo {volume} {7}},\
  \bibinfo {pages} {13017} (\bibinfo {year} {2016})}\BibitemShut {NoStop}%
\bibitem [{\citenamefont {Shamblin}\ \emph {et~al.}(2018)\citenamefont
  {Shamblin}, \citenamefont {Heres}, \citenamefont {Zhou}, \citenamefont
  {Sangoro}, \citenamefont {Lang}, \citenamefont {Neuefeind}, \citenamefont
  {Alonso},\ and\ \citenamefont {Johnston}}]{Shamblin:2018p86}%
  \BibitemOpen
  \bibfield  {author} {\bibinfo {author} {\bibfnamefont {J.}~\bibnamefont
  {Shamblin}}, \bibinfo {author} {\bibfnamefont {M.}~\bibnamefont {Heres}},
  \bibinfo {author} {\bibfnamefont {H.}~\bibnamefont {Zhou}}, \bibinfo {author}
  {\bibfnamefont {J.}~\bibnamefont {Sangoro}}, \bibinfo {author} {\bibfnamefont
  {M.}~\bibnamefont {Lang}}, \bibinfo {author} {\bibfnamefont {J.}~\bibnamefont
  {Neuefeind}}, \bibinfo {author} {\bibfnamefont {J.~A.}\ \bibnamefont
  {Alonso}}, \ and\ \bibinfo {author} {\bibfnamefont {S.}~\bibnamefont
  {Johnston}},\ }\href {\doibase 10.1038/s41467-017-02561-6} {\bibfield
  {journal} {\bibinfo  {journal} {Nature Communications}\ }\textbf {\bibinfo
  {volume} {9}},\ \bibinfo {pages} {86} (\bibinfo {year} {2018})}\BibitemShut
  {NoStop}%
\bibitem [{\citenamefont {Lee}\ \emph {et~al.}(2011{\natexlab{a}})\citenamefont
  {Lee}, \citenamefont {Chen},\ and\ \citenamefont
  {Balents}}]{Lee:2011p016405}%
  \BibitemOpen
  \bibfield  {author} {\bibinfo {author} {\bibfnamefont {S.}~\bibnamefont
  {Lee}}, \bibinfo {author} {\bibfnamefont {R.}~\bibnamefont {Chen}}, \ and\
  \bibinfo {author} {\bibfnamefont {L.}~\bibnamefont {Balents}},\ }\href
  {\doibase 10.1103/PhysRevLett.106.016405} {\bibfield  {journal} {\bibinfo
  {journal} {Phys. Rev. Lett.}\ }\textbf {\bibinfo {volume} {106}},\ \bibinfo
  {pages} {016405} (\bibinfo {year} {2011}{\natexlab{a}})}\BibitemShut
  {NoStop}%
\bibitem [{\citenamefont {Lee}\ \emph {et~al.}(2011{\natexlab{b}})\citenamefont
  {Lee}, \citenamefont {Chen},\ and\ \citenamefont
  {Balents}}]{Lee:2011p165119}%
  \BibitemOpen
  \bibfield  {author} {\bibinfo {author} {\bibfnamefont {S.}~\bibnamefont
  {Lee}}, \bibinfo {author} {\bibfnamefont {R.}~\bibnamefont {Chen}}, \ and\
  \bibinfo {author} {\bibfnamefont {L.}~\bibnamefont {Balents}},\ }\href
  {\doibase 10.1103/PhysRevB.84.165119} {\bibfield  {journal} {\bibinfo
  {journal} {Phys. Rev. B}\ }\textbf {\bibinfo {volume} {84}},\ \bibinfo
  {pages} {165119} (\bibinfo {year} {2011}{\natexlab{b}})}\BibitemShut
  {NoStop}%
\bibitem [{\citenamefont {Hepting}\ \emph {et~al.}(2014)\citenamefont
  {Hepting}, \citenamefont {Minola}, \citenamefont {Frano}, \citenamefont
  {Cristiani}, \citenamefont {Logvenov}, \citenamefont {Schierle},
  \citenamefont {Wu}, \citenamefont {Bluschke}, \citenamefont {Weschke},
  \citenamefont {Habermeier}, \citenamefont {Benckiser}, \citenamefont
  {Le~Tacon},\ and\ \citenamefont {Keimer}}]{Hepting:2014p227206}%
  \BibitemOpen
  \bibfield  {author} {\bibinfo {author} {\bibfnamefont {M.}~\bibnamefont
  {Hepting}}, \bibinfo {author} {\bibfnamefont {M.}~\bibnamefont {Minola}},
  \bibinfo {author} {\bibfnamefont {A.}~\bibnamefont {Frano}}, \bibinfo
  {author} {\bibfnamefont {G.}~\bibnamefont {Cristiani}}, \bibinfo {author}
  {\bibfnamefont {G.}~\bibnamefont {Logvenov}}, \bibinfo {author}
  {\bibfnamefont {E.}~\bibnamefont {Schierle}}, \bibinfo {author}
  {\bibfnamefont {M.}~\bibnamefont {Wu}}, \bibinfo {author} {\bibfnamefont
  {M.}~\bibnamefont {Bluschke}}, \bibinfo {author} {\bibfnamefont
  {E.}~\bibnamefont {Weschke}}, \bibinfo {author} {\bibfnamefont {H.-U.}\
  \bibnamefont {Habermeier}}, \bibinfo {author} {\bibfnamefont
  {E.}~\bibnamefont {Benckiser}}, \bibinfo {author} {\bibfnamefont
  {M.}~\bibnamefont {Le~Tacon}}, \ and\ \bibinfo {author} {\bibfnamefont
  {B.}~\bibnamefont {Keimer}},\ }\href {\doibase
  10.1103/PhysRevLett.113.227206} {\bibfield  {journal} {\bibinfo  {journal}
  {Phys. Rev. Lett.}\ }\textbf {\bibinfo {volume} {113}},\ \bibinfo {pages}
  {227206} (\bibinfo {year} {2014})}\BibitemShut {NoStop}%
\bibitem [{\citenamefont {Middey}\ \emph
  {et~al.}(2018{\natexlab{a}})\citenamefont {Middey}, \citenamefont {Meyers},
  \citenamefont {Kareev}, \citenamefont {Cao}, \citenamefont {Liu},
  \citenamefont {Shafer}, \citenamefont {Freeland}, \citenamefont {Kim},
  \citenamefont {Ryan},\ and\ \citenamefont {Chakhalian}}]{Middey:2018p156801}%
  \BibitemOpen
  \bibfield  {author} {\bibinfo {author} {\bibfnamefont {S.}~\bibnamefont
  {Middey}}, \bibinfo {author} {\bibfnamefont {D.}~\bibnamefont {Meyers}},
  \bibinfo {author} {\bibfnamefont {M.}~\bibnamefont {Kareev}}, \bibinfo
  {author} {\bibfnamefont {Y.}~\bibnamefont {Cao}}, \bibinfo {author}
  {\bibfnamefont {X.}~\bibnamefont {Liu}}, \bibinfo {author} {\bibfnamefont
  {P.}~\bibnamefont {Shafer}}, \bibinfo {author} {\bibfnamefont {J.~W.}\
  \bibnamefont {Freeland}}, \bibinfo {author} {\bibfnamefont {J.-W.}\
  \bibnamefont {Kim}}, \bibinfo {author} {\bibfnamefont {P.~J.}\ \bibnamefont
  {Ryan}}, \ and\ \bibinfo {author} {\bibfnamefont {J.}~\bibnamefont
  {Chakhalian}},\ }\href {\doibase 10.1103/PhysRevLett.120.156801} {\bibfield
  {journal} {\bibinfo  {journal} {Phys. Rev. Lett.}\ }\textbf {\bibinfo
  {volume} {120}},\ \bibinfo {pages} {156801} (\bibinfo {year}
  {2018}{\natexlab{a}})}\BibitemShut {NoStop}%
\bibitem{nno_fet}R. Scherwitzl,  P. Zubko,  I. Gutierrez Lezama,  S. Ono,  A. F. Morpurgo,  G. Catalan,  and J.?M. Triscone, Advanced Materials, {\bf 22}, 5517 (2010).
\bibitem{Shi2013} J . Shi , S. D. Ha , Y. Zhou, F. Schoofs, and S. Ramanathan Nat. Commun. {\bf 4}, 2676 (2013).
\bibitem{Shi2014} J. Shi, Y. Zhou and S. Ramanathan, Nat. Commun. {\bf 5}, 4860 (2014).
\bibitem{Ha2014} S. D. Ha , J. Shi, Y. Meroz, L. Mahadevan,  and S. Ramanathan, Phys. Rev. Appl. {\bf 2}, 064003 (2014).
\bibitem{noisedevice}B. K. Jones - IEE Proceedings-Circuits, Devices and Systems, {\bf 149}, 13 (2002).
\bibitem [{\citenamefont {Koushik}\ \emph {et~al.}(2011)\citenamefont
  {Koushik}, \citenamefont {Baenninger}, \citenamefont {Narayan}, \citenamefont
  {Mukerjee}, \citenamefont {Pepper}, \citenamefont {Farrer}, \citenamefont
  {Ritchie},\ and\ \citenamefont {Ghosh}}]{koushik2011evidence}%
  \BibitemOpen
  \bibfield  {author} {\bibinfo {author} {\bibfnamefont {R.}~\bibnamefont
  {Koushik}}, \bibinfo {author} {\bibfnamefont {M.}~\bibnamefont {Baenninger}},
  \bibinfo {author} {\bibfnamefont {V.}~\bibnamefont {Narayan}}, \bibinfo
  {author} {\bibfnamefont {S.}~\bibnamefont {Mukerjee}}, \bibinfo {author}
  {\bibfnamefont {M.}~\bibnamefont {Pepper}}, \bibinfo {author} {\bibfnamefont
  {I.}~\bibnamefont {Farrer}}, \bibinfo {author} {\bibfnamefont {D.~A.}\
  \bibnamefont {Ritchie}}, \ and\ \bibinfo {author} {\bibfnamefont
  {A.}~\bibnamefont {Ghosh}},\ }\href@noop {} {\bibfield  {journal} {\bibinfo
  {journal} {Physical Review B}\ }\textbf {\bibinfo {volume} {83}},\ \bibinfo
  {pages} {085302} (\bibinfo {year} {2011})}\BibitemShut {NoStop}%
\bibitem [{\citenamefont {Chandni}\ \emph {et~al.}(2009)\citenamefont
  {Chandni}, \citenamefont {Ghosh}, \citenamefont {Vijaya},\ and\ \citenamefont
  {Mohan}}]{chandni2009criticality}%
  \BibitemOpen
  \bibfield  {author} {\bibinfo {author} {\bibfnamefont {U.}~\bibnamefont
  {Chandni}}, \bibinfo {author} {\bibfnamefont {A.}~\bibnamefont {Ghosh}},
  \bibinfo {author} {\bibfnamefont {H.}~\bibnamefont {Vijaya}}, \ and\ \bibinfo
  {author} {\bibfnamefont {S.}~\bibnamefont {Mohan}},\ }\href@noop {}
  {\bibfield  {journal} {\bibinfo  {journal} {Physical review letters}\
  }\textbf {\bibinfo {volume} {102}},\ \bibinfo {pages} {025701} (\bibinfo
  {year} {2009})}\BibitemShut {NoStop}%
\bibitem [{\citenamefont {Kundu}\ \emph {et~al.}(2017)\citenamefont {Kundu},
  \citenamefont {Ray}, \citenamefont {Dolui}, \citenamefont {Bagwe},
  \citenamefont {Choudhury}, \citenamefont {Krupanidhi}, \citenamefont {Das},
  \citenamefont {Raychaudhuri},\ and\ \citenamefont {Bid}}]{kundu2017quantum}%
  \BibitemOpen
  \bibfield  {author} {\bibinfo {author} {\bibfnamefont {H.~K.}\ \bibnamefont
  {Kundu}}, \bibinfo {author} {\bibfnamefont {S.}~\bibnamefont {Ray}}, \bibinfo
  {author} {\bibfnamefont {K.}~\bibnamefont {Dolui}}, \bibinfo {author}
  {\bibfnamefont {V.}~\bibnamefont {Bagwe}}, \bibinfo {author} {\bibfnamefont
  {P.~R.}\ \bibnamefont {Choudhury}}, \bibinfo {author} {\bibfnamefont
  {S.}~\bibnamefont {Krupanidhi}}, \bibinfo {author} {\bibfnamefont
  {T.}~\bibnamefont {Das}}, \bibinfo {author} {\bibfnamefont {P.}~\bibnamefont
  {Raychaudhuri}}, \ and\ \bibinfo {author} {\bibfnamefont {A.}~\bibnamefont
  {Bid}},\ }\href@noop {} {\bibfield  {journal} {\bibinfo  {journal} {Physical
  review letters}\ }\textbf {\bibinfo {volume} {119}},\ \bibinfo {pages}
  {226802} (\bibinfo {year} {2017})}\BibitemShut {NoStop}%
\bibitem [{\citenamefont {Clarke}\ and\ \citenamefont
  {Hsiang}(1976)}]{clarke1976low}%
  \BibitemOpen
  \bibfield  {author} {\bibinfo {author} {\bibfnamefont {J.}~\bibnamefont
  {Clarke}}\ and\ \bibinfo {author} {\bibfnamefont {T.~Y.}\ \bibnamefont
  {Hsiang}},\ }\href@noop {} {\bibfield  {journal} {\bibinfo  {journal}
  {Physical Review B}\ }\textbf {\bibinfo {volume} {13}},\ \bibinfo {pages}
  {4790} (\bibinfo {year} {1976})}\BibitemShut {NoStop}%
\bibitem [{\citenamefont {Babi{\'c}}\ \emph {et~al.}(2007)\citenamefont
  {Babi{\'c}}, \citenamefont {Bentner}, \citenamefont {S{\"u}rgers},\ and\
  \citenamefont {Strunk}}]{babic20071}%
  \BibitemOpen
  \bibfield  {author} {\bibinfo {author} {\bibfnamefont {D.}~\bibnamefont
  {Babi{\'c}}}, \bibinfo {author} {\bibfnamefont {J.}~\bibnamefont {Bentner}},
  \bibinfo {author} {\bibfnamefont {C.}~\bibnamefont {S{\"u}rgers}}, \ and\
  \bibinfo {author} {\bibfnamefont {C.}~\bibnamefont {Strunk}},\ }\href@noop {}
  {\bibfield  {journal} {\bibinfo  {journal} {Physical Review B}\ }\textbf
  {\bibinfo {volume} {76}},\ \bibinfo {pages} {134515} (\bibinfo {year}
  {2007})}\BibitemShut {NoStop}%
\bibitem{reif2009fundamentals}F. Reif, Fundamentals of statistical and thermal physics, Waveland Press  (2009).
\bibitem [{\citenamefont {Weissman}(1988)}]{RevModPhys.60.537}%
  \BibitemOpen
  \bibfield  {author} {\bibinfo {author} {\bibfnamefont {M.~B.}\ \bibnamefont
  {Weissman}},\ }\href@noop {} {\bibfield  {journal} {\bibinfo  {journal} {Rev.
  Mod. Phys.}\ }\textbf {\bibinfo {volume} {60}},\ \bibinfo {pages} {537}
  (\bibinfo {year} {1988})}\BibitemShut {NoStop}%
\bibitem [{\citenamefont {Ghosh}\ \emph {et~al.}(2004)\citenamefont {Ghosh},
  \citenamefont {Kar}, \citenamefont {Bid},\ and\ \citenamefont
  {Raychaudhuri}}]{ghosh2004set}%
  \BibitemOpen
  \bibfield  {author} {\bibinfo {author} {\bibfnamefont {A.}~\bibnamefont
  {Ghosh}}, \bibinfo {author} {\bibfnamefont {S.}~\bibnamefont {Kar}}, \bibinfo
  {author} {\bibfnamefont {A.}~\bibnamefont {Bid}}, \ and\ \bibinfo {author}
  {\bibfnamefont {A.}~\bibnamefont {Raychaudhuri}},\ }\href@noop {} {\bibfield
  {journal} {\bibinfo  {journal} {arXiv preprint cond-mat/0402130}\ } (\bibinfo
  {year} {2004})}\BibitemShut {NoStop}%
\bibitem [{\citenamefont {Sahoo}\ \emph {et~al.}(2014)\citenamefont {Sahoo},
  \citenamefont {Ha}, \citenamefont {Ramanathan},\ and\ \citenamefont
  {Ghosh}}]{sahoo2014conductivity}%
  \BibitemOpen
  \bibfield  {author} {\bibinfo {author} {\bibfnamefont {A.}~\bibnamefont
  {Sahoo}}, \bibinfo {author} {\bibfnamefont {S.~D.}\ \bibnamefont {Ha}},
  \bibinfo {author} {\bibfnamefont {S.}~\bibnamefont {Ramanathan}}, \ and\
  \bibinfo {author} {\bibfnamefont {A.}~\bibnamefont {Ghosh}},\ }\href@noop {}
  {\bibfield  {journal} {\bibinfo  {journal} {Physical Review B}\ }\textbf
  {\bibinfo {volume} {90}},\ \bibinfo {pages} {085116} (\bibinfo {year}
  {2014})}\BibitemShut {NoStop}%
\bibitem [{\citenamefont {Alsaqqa}\ \emph {et~al.}(2017)\citenamefont
  {Alsaqqa}, \citenamefont {Singh}, \citenamefont {Middey}, \citenamefont
  {Kareev}, \citenamefont {Chakhalian},\ and\ \citenamefont
  {Sambandamurthy}}]{alsaqqa2017phase}%
  \BibitemOpen
  \bibfield  {author} {\bibinfo {author} {\bibfnamefont {A.~M.}\ \bibnamefont
  {Alsaqqa}}, \bibinfo {author} {\bibfnamefont {S.}~\bibnamefont {Singh}},
  \bibinfo {author} {\bibfnamefont {S.}~\bibnamefont {Middey}}, \bibinfo
  {author} {\bibfnamefont {M.}~\bibnamefont {Kareev}}, \bibinfo {author}
  {\bibfnamefont {J.}~\bibnamefont {Chakhalian}}, \ and\ \bibinfo {author}
  {\bibfnamefont {G.}~\bibnamefont {Sambandamurthy}},\ }\href@noop {}
  {\bibfield  {journal} {\bibinfo  {journal} {Physical Review B}\ }\textbf
  {\bibinfo {volume} {95}},\ \bibinfo {pages} {125132} (\bibinfo {year}
  {2017})}\BibitemShut {NoStop}%
\bibitem [{\citenamefont {Bisht}\ \emph {et~al.}(2017)\citenamefont {Bisht},
  \citenamefont {Samanta},\ and\ \citenamefont
  {Raychaudhuri}}]{Bisht:2017p115147}%
  \BibitemOpen
  \bibfield  {author} {\bibinfo {author} {\bibfnamefont {R.~S.}\ \bibnamefont
  {Bisht}}, \bibinfo {author} {\bibfnamefont {S.}~\bibnamefont {Samanta}}, \
  and\ \bibinfo {author} {\bibfnamefont {A.~K.}\ \bibnamefont {Raychaudhuri}},\
  }\href {\doibase 10.1103/PhysRevB.95.115147} {\bibfield  {journal} {\bibinfo
  {journal} {Phys. Rev. B}\ }\textbf {\bibinfo {volume} {95}},\ \bibinfo
  {pages} {115147} (\bibinfo {year} {2017})}\BibitemShut {NoStop}%
\bibitem [{\citenamefont {Bid}\ \emph {et~al.}(2003)\citenamefont {Bid},
  \citenamefont {Guha},\ and\ \citenamefont {Raychaudhuri}}]{bid2003low}%
  \BibitemOpen
  \bibfield  {author} {\bibinfo {author} {\bibfnamefont {A.}~\bibnamefont
  {Bid}}, \bibinfo {author} {\bibfnamefont {A.}~\bibnamefont {Guha}}, \ and\
  \bibinfo {author} {\bibfnamefont {A.}~\bibnamefont {Raychaudhuri}},\
  }\href@noop {} {\bibfield  {journal} {\bibinfo  {journal} {Physical Review
  B}\ }\textbf {\bibinfo {volume} {67}},\ \bibinfo {pages} {174415} (\bibinfo
  {year} {2003})}\BibitemShut {NoStop}%
\bibitem [{\citenamefont {Middey}\ \emph
  {et~al.}(2018{\natexlab{b}})\citenamefont {Middey}, \citenamefont {Meyers},
  \citenamefont {Kareev}, \citenamefont {Liu}, \citenamefont {Cao},
  \citenamefont {Freeland},\ and\ \citenamefont
  {Chakhalian}}]{Middey:2018p045115}%
  \BibitemOpen
  \bibfield  {author} {\bibinfo {author} {\bibfnamefont {S.}~\bibnamefont
  {Middey}}, \bibinfo {author} {\bibfnamefont {D.}~\bibnamefont {Meyers}},
  \bibinfo {author} {\bibfnamefont {M.}~\bibnamefont {Kareev}}, \bibinfo
  {author} {\bibfnamefont {X.}~\bibnamefont {Liu}}, \bibinfo {author}
  {\bibfnamefont {Y.}~\bibnamefont {Cao}}, \bibinfo {author} {\bibfnamefont
  {J.~W.}\ \bibnamefont {Freeland}}, \ and\ \bibinfo {author} {\bibfnamefont
  {J.}~\bibnamefont {Chakhalian}},\ }\href@noop {} {\bibfield  {journal}
  {\bibinfo  {journal} {Phys. Rev. B}\ }\textbf {\bibinfo {volume} {98}},\
  \bibinfo {pages} {045115} (\bibinfo {year} {2018}{\natexlab{b}})}\BibitemShut
  {NoStop}%
\bibitem [{\citenamefont {Middey}\ \emph
{et~al.}(2018{\natexlab{c}})\citenamefont {Middey}, \citenamefont {Meyers},
\citenamefont {Patel}, \citenamefont {Liu}, \citenamefont {Kareev},
\citenamefont {Shafer}, \citenamefont {Kim}, \citenamefont {Ryan},\ and\
\citenamefont {Chakhalian}}]{Middey:2018apl}%
\BibitemOpen
 \bibfield  {author} {\bibinfo {author} {\bibfnamefont {S.}~\bibnamefont
{Middey}}, \bibinfo {author} {\bibfnamefont {D.}~\bibnamefont {Meyers}},
 \bibinfo {author} {\bibfnamefont {R.~K.}\ \bibnamefont {Patel}}, \bibinfo
{author} {\bibfnamefont {X.}~\bibnamefont {Liu}}, \bibinfo {author}
{\bibfnamefont {M.}~\bibnamefont {Kareev}}, \bibinfo {author} {\bibfnamefont
{P.}~\bibnamefont {Shafer}}, \bibinfo {author} {\bibfnamefont {J.-W.}\
\bibnamefont {Kim}}, \bibinfo {author} {\bibfnamefont {P.~J.}\ \bibnamefont
{Ryan}}, \ and\ \bibinfo {author} {\bibfnamefont {J.}~\bibnamefont
{Chakhalian}},\ }\href@noop {} {\bibfield  {journal} {\bibinfo  {journal}
{Applied Physics Letters}\ }\textbf {\bibinfo {volume} {113}},\ \bibinfo
{pages} {081602} (\bibinfo {year} {2018}{\natexlab{c}})}\BibitemShut {NoStop}%
\bibitem [{\citenamefont {Zhou}, \citenamefont {Goodenough},\ and\
  \citenamefont {Dabrowski}(2005)}]{Zhou:2005p226602}%
  \BibitemOpen
  \bibfield  {author} {\bibinfo {author} {\bibfnamefont {J.-S.}\ \bibnamefont
  {Zhou}}, \bibinfo {author} {\bibfnamefont {J.~B.}\ \bibnamefont
  {Goodenough}}, \ and\ \bibinfo {author} {\bibfnamefont {B.}~\bibnamefont
  {Dabrowski}},\ }\href {\doibase 10.1103/PhysRevLett.94.226602} {\bibfield
  {journal} {\bibinfo  {journal} {Phys. Rev. Lett.}\ }\textbf {\bibinfo
  {volume} {94}},\ \bibinfo {pages} {226602} (\bibinfo {year}
  {2005})}\BibitemShut {NoStop}%
  \bibitem{Ojha2019} S. K. Ojha, S. Ray, T. Das, S. Middey, S. Sarkar, P. Mahadevan, Z. Wang, Y. Zhu, X. Liu, M. Kareev, and J. Chakhalian, Phys. Rev. B {\bf 99},  235153 (2019).
\bibitem [{\citenamefont {Hooge}(1969)}]{hooge19691}%
  \BibitemOpen
  \bibfield  {author} {\bibinfo {author} {\bibfnamefont {F.~N.}\ \bibnamefont
  {Hooge}},\ }\href@noop {} {\bibfield  {journal} {\bibinfo  {journal} {Phys.
  Lett.}\ }\textbf {\bibinfo {volume} {29}},\ \bibinfo {pages} {139} (\bibinfo
  {year} {1969})}\BibitemShut {NoStop}%
\bibitem [{\citenamefont {Piamonteze}\ \emph {et~al.}(2005)\citenamefont
  {Piamonteze}, \citenamefont {Tolentino}, \citenamefont {Ramos}, \citenamefont
  {Massa}, \citenamefont {Alonso}, \citenamefont {Mart\'{i}nez-Lope},\ and\
  \citenamefont {Casais}}]{Piamonteze:2005p012104}%
  \BibitemOpen
  \bibfield  {author} {\bibinfo {author} {\bibfnamefont {C.}~\bibnamefont
  {Piamonteze}}, \bibinfo {author} {\bibfnamefont {H.~C.~N.}\ \bibnamefont
  {Tolentino}}, \bibinfo {author} {\bibfnamefont {A.~Y.}\ \bibnamefont
  {Ramos}}, \bibinfo {author} {\bibfnamefont {N.~E.}\ \bibnamefont {Massa}},
  \bibinfo {author} {\bibfnamefont {J.~A.}\ \bibnamefont {Alonso}}, \bibinfo
  {author} {\bibfnamefont {M.~J.}\ \bibnamefont {Mart\'{i}nez-Lope}}, \ and\
  \bibinfo {author} {\bibfnamefont {M.~T.}\ \bibnamefont {Casais}},\ }\href
  {\doibase 10.1103/PhysRevB.71.012104} {\bibfield  {journal} {\bibinfo
  {journal} {Phys. Rev. B}\ }\textbf {\bibinfo {volume} {71}},\ \bibinfo
  {pages} {012104} (\bibinfo {year} {2005})}\BibitemShut {NoStop}%
\bibitem{hooge1972discussion} F. N. Hooge, Physica {\bf 60}, 130 (1972).
\bibitem [{\citenamefont {Tremblay}\ \emph {et~al.}(1986)\citenamefont
  {Tremblay}, \citenamefont {Feng},\ and\ \citenamefont
  {Breton}}]{PhysRevB.33.2077}%
  \BibitemOpen
  \bibfield  {author} {\bibinfo {author} {\bibfnamefont {A.-M.~S.}\
  \bibnamefont {Tremblay}}, \bibinfo {author} {\bibfnamefont {S.}~\bibnamefont
  {Feng}}, \ and\ \bibinfo {author} {\bibfnamefont {P.}~\bibnamefont
  {Breton}},\ }\href@noop {} {\bibfield  {journal} {\bibinfo  {journal} {Phys.
  Rev. B}\ }\textbf {\bibinfo {volume} {33}},\ \bibinfo {pages} {2077}
  (\bibinfo {year} {1986})}\BibitemShut {NoStop}%
\bibitem [{\citenamefont {Kiss}\ and\ \citenamefont
  {Svedlindh}(1994)}]{kiss1994noise}%
  \BibitemOpen
  \bibfield  {author} {\bibinfo {author} {\bibfnamefont {L.~B.}\ \bibnamefont
  {Kiss}}\ and\ \bibinfo {author} {\bibfnamefont {P.}~\bibnamefont
  {Svedlindh}},\ }\href@noop {} {\bibfield  {journal} {\bibinfo  {journal}
  {IEEE Transactions on Electron Devices}\ }\textbf {\bibinfo {volume} {41}},\
  \bibinfo {pages} {2112} (\bibinfo {year} {1994})}\BibitemShut {NoStop}%
\bibitem [{\citenamefont {Daptary}\ \emph {et~al.}(2016)\citenamefont
  {Daptary}, \citenamefont {Kumar}, \citenamefont {Kumar}, \citenamefont
  {Dogra}, \citenamefont {Mohanta}, \citenamefont {Taraphder},\ and\
  \citenamefont {Bid}}]{daptary2016correlated}%
  \BibitemOpen
  \bibfield  {author} {\bibinfo {author} {\bibfnamefont {G.~N.}\ \bibnamefont
  {Daptary}}, \bibinfo {author} {\bibfnamefont {S.}~\bibnamefont {Kumar}},
  \bibinfo {author} {\bibfnamefont {P.}~\bibnamefont {Kumar}}, \bibinfo
  {author} {\bibfnamefont {A.}~\bibnamefont {Dogra}}, \bibinfo {author}
  {\bibfnamefont {N.}~\bibnamefont {Mohanta}}, \bibinfo {author} {\bibfnamefont
  {A.}~\bibnamefont {Taraphder}}, \ and\ \bibinfo {author} {\bibfnamefont
  {A.}~\bibnamefont {Bid}},\ }\href@noop {} {\bibfield  {journal} {\bibinfo
  {journal} {Physical Review B}\ }\textbf {\bibinfo {volume} {94}},\ \bibinfo
  {pages} {085104} (\bibinfo {year} {2016})}\BibitemShut {NoStop}%
\bibitem [{\citenamefont {Dutta}\ and\ \citenamefont
  {Horn}(1981)}]{dutta1981low}%
  \BibitemOpen
  \bibfield  {author} {\bibinfo {author} {\bibfnamefont {P.}~\bibnamefont
  {Dutta}}\ and\ \bibinfo {author} {\bibfnamefont {P.}~\bibnamefont {Horn}},\
  }\href@noop {} {\bibfield  {journal} {\bibinfo  {journal} {Reviews of Modern
  physics}\ }\textbf {\bibinfo {volume} {53}},\ \bibinfo {pages} {497}
  (\bibinfo {year} {1981})}\BibitemShut {NoStop}%
\bibitem [{\citenamefont {Daptary}\ \emph {et~al.}(2017)\citenamefont
  {Daptary}, \citenamefont {Sow}, \citenamefont {Sarkar}, \citenamefont
  {Chiniwar}, \citenamefont {Kumar}, \citenamefont {Sil},\ and\ \citenamefont
  {Bid}}]{daptary2017effect}%
  \BibitemOpen
  \bibfield  {author} {\bibinfo {author} {\bibfnamefont {G.~N.}\ \bibnamefont
  {Daptary}}, \bibinfo {author} {\bibfnamefont {C.}~\bibnamefont {Sow}},
  \bibinfo {author} {\bibfnamefont {S.}~\bibnamefont {Sarkar}}, \bibinfo
  {author} {\bibfnamefont {S.}~\bibnamefont {Chiniwar}}, \bibinfo {author}
  {\bibfnamefont {P.~A.}\ \bibnamefont {Kumar}}, \bibinfo {author}
  {\bibfnamefont {A.}~\bibnamefont {Sil}}, \ and\ \bibinfo {author}
  {\bibfnamefont {A.}~\bibnamefont {Bid}},\ }\href@noop {} {\bibfield
  {journal} {\bibinfo  {journal} {Physica B: Condensed Matter}\ }\textbf
  {\bibinfo {volume} {511}},\ \bibinfo {pages} {74} (\bibinfo {year}
  {2017})}\BibitemShut {NoStop}%
\bibitem [{\citenamefont {Seidler}\ \emph {et~al.}(1996)\citenamefont
  {Seidler}, \citenamefont {Solin},\ and\ \citenamefont
  {Marley}}]{seidler1996dynamical}%
  \BibitemOpen
  \bibfield  {author} {\bibinfo {author} {\bibfnamefont {G.}~\bibnamefont
  {Seidler}}, \bibinfo {author} {\bibfnamefont {S.}~\bibnamefont {Solin}}, \
  and\ \bibinfo {author} {\bibfnamefont {A.}~\bibnamefont {Marley}},\
  }\href@noop {} {\bibfield  {journal} {\bibinfo  {journal} {Physical review
  letters}\ }\textbf {\bibinfo {volume} {76}},\ \bibinfo {pages} {3049}
  (\bibinfo {year} {1996})}\BibitemShut {NoStop}%
\bibitem [{\citenamefont {Daptary}\ \emph {et~al.}(2014)\citenamefont
  {Daptary}, \citenamefont {Sow}, \citenamefont {Kumar},\ and\ \citenamefont
  {Bid}}]{daptary2014probing}%
  \BibitemOpen
  \bibfield  {author} {\bibinfo {author} {\bibfnamefont {G.~N.}\ \bibnamefont
  {Daptary}}, \bibinfo {author} {\bibfnamefont {C.}~\bibnamefont {Sow}},
  \bibinfo {author} {\bibfnamefont {P.~A.}\ \bibnamefont {Kumar}}, \ and\
  \bibinfo {author} {\bibfnamefont {A.}~\bibnamefont {Bid}},\ }\href@noop {}
  {\bibfield  {journal} {\bibinfo  {journal} {Physical Review B}\ }\textbf
  {\bibinfo {volume} {90}},\ \bibinfo {pages} {115153} (\bibinfo {year}
  {2014})}\BibitemShut {NoStop}%
\bibitem [{\citenamefont {Seidler}\ and\ \citenamefont
  {Solin}(1996)}]{seidler1996non}%
  \BibitemOpen
  \bibfield  {author} {\bibinfo {author} {\bibfnamefont {G.}~\bibnamefont
  {Seidler}}\ and\ \bibinfo {author} {\bibfnamefont {S.}~\bibnamefont
  {Solin}},\ }\href@noop {} {\bibfield  {journal} {\bibinfo  {journal}
  {Physical Review B}\ }\textbf {\bibinfo {volume} {53}},\ \bibinfo {pages}
  {9753} (\bibinfo {year} {1996})}\BibitemShut {NoStop}%
\bibitem [{\citenamefont {Eguchi}\ \emph {et~al.}(2009)\citenamefont {Eguchi},
  \citenamefont {Chainani}, \citenamefont {Taguchi}, \citenamefont {Matsunami},
  \citenamefont {Ishida}, \citenamefont {Horiba}, \citenamefont {Senba},
  \citenamefont {Ohashi},\ and\ \citenamefont {Shin}}]{Eguchi:2009p115122}%
  \BibitemOpen
  \bibfield  {author} {\bibinfo {author} {\bibfnamefont {R.}~\bibnamefont
  {Eguchi}}, \bibinfo {author} {\bibfnamefont {A.}~\bibnamefont {Chainani}},
  \bibinfo {author} {\bibfnamefont {M.}~\bibnamefont {Taguchi}}, \bibinfo
  {author} {\bibfnamefont {M.}~\bibnamefont {Matsunami}}, \bibinfo {author}
  {\bibfnamefont {Y.}~\bibnamefont {Ishida}}, \bibinfo {author} {\bibfnamefont
  {K.}~\bibnamefont {Horiba}}, \bibinfo {author} {\bibfnamefont
  {Y.}~\bibnamefont {Senba}}, \bibinfo {author} {\bibfnamefont
  {H.}~\bibnamefont {Ohashi}}, \ and\ \bibinfo {author} {\bibfnamefont
  {S.}~\bibnamefont {Shin}},\ }\href {\doibase 10.1103/PhysRevB.79.115122}
  {\bibfield  {journal} {\bibinfo  {journal} {Phys. Rev. B}\ }\textbf {\bibinfo
  {volume} {79}},\ \bibinfo {pages} {115122} (\bibinfo {year}
  {2009})}\BibitemShut {NoStop}%
\bibitem [{\citenamefont {Zhou}\ \emph {et~al.}(2004)\citenamefont {Zhou},
  \citenamefont {Goodenough},\ and\ \citenamefont
  {Dabrowski}}]{Zhou:2004p081102}%
  \BibitemOpen
  \bibfield  {author} {\bibinfo {author} {\bibfnamefont {J.-S.}\ \bibnamefont
  {Zhou}}, \bibinfo {author} {\bibfnamefont {J.~B.}\ \bibnamefont
  {Goodenough}}, \ and\ \bibinfo {author} {\bibfnamefont {B.}~\bibnamefont
  {Dabrowski}},\ }\href {\doibase 10.1103/PhysRevB.70.081102} {\bibfield
  {journal} {\bibinfo  {journal} {Phys. Rev. B}\ }\textbf {\bibinfo {volume}
  {70}},\ \bibinfo {pages} {081102} (\bibinfo {year} {2004})}\BibitemShut
  {NoStop}%
\bibitem [{\citenamefont {Preziosi}\ \emph {et~al.}(2018)\citenamefont
  {Preziosi}, \citenamefont {Lopez-Mir}, \citenamefont {Li}, \citenamefont
  {Cornelissen}, \citenamefont {Lee}, \citenamefont {Trier}, \citenamefont
  {Bouzehouane}, \citenamefont {Valencia}, \citenamefont {Gloter},
  \citenamefont {Barthélémy} \emph {et~al.}}]{preziosi:2018p2226}%
  \BibitemOpen
  \bibfield  {author} {\bibinfo {author} {\bibfnamefont {D.}~\bibnamefont
  {Preziosi}}, \bibinfo {author} {\bibfnamefont {L.}~\bibnamefont {Lopez-Mir}},
  \bibinfo {author} {\bibfnamefont {X.}~\bibnamefont {Li}}, \bibinfo {author}
  {\bibfnamefont {T.}~\bibnamefont {Cornelissen}}, \bibinfo {author}
  {\bibfnamefont {J.~H.}\ \bibnamefont {Lee}}, \bibinfo {author} {\bibfnamefont
  {F.}~\bibnamefont {Trier}}, \bibinfo {author} {\bibfnamefont
  {K.}~\bibnamefont {Bouzehouane}}, \bibinfo {author} {\bibfnamefont
  {S.}~\bibnamefont {Valencia}}, \bibinfo {author} {\bibfnamefont
  {A.}~\bibnamefont {Gloter}}, \bibinfo {author} {\bibfnamefont
  {A.}~\bibnamefont {Barthélémy}},  \emph {et~al.},\ }\href@noop {} {\bibfield
  {journal} {\bibinfo  {journal} {Nano letters}\ }\textbf {\bibinfo {volume}
  {18}},\ \bibinfo {pages} {2226} (\bibinfo {year} {2018})}\BibitemShut
  {NoStop}%
  \bibitem{ghosh1997dependence} A. Ghosh, A. K. Raychaudhuri, R. Sreekala, M. Rajeswari, and T. Venkatesan, Journal of Physics D: Applied Physics, {\bf 30}, L75 (1997).
\end{thebibliography}
%

\end{document}